\documentclass[acmsmall, natbib, screen]{acmart}

\acmSubmissionID{}

\usepackage{multirow}
\usepackage{mathtools}
\usepackage{svg}
\usepackage[font=small,skip=0pt]{caption}
\usepackage{graphicx}
\usepackage{subcaption}
\usepackage[inline]{enumitem}
\usepackage{booktabs}
\usepackage{adjustbox}
\usepackage{hyphenat}
\usepackage{acronym}
\usepackage{cleveref}[2012/02/15]
\crefformat{footnote}{#2\footnotemark[#1]#3}
\usepackage{pifont}
\newcommand{\cmark}{\ding{51}}%

\usepackage{bigdelim}

\makeatletter
\renewcommand\paragraph{\@startsection{paragraph}{4}{\z@}%
                       {-2\p@ \@plus -1\p@ \@minus -1\p@}%
                       {-0.5em \@plus -0.22em \@minus -0.1em}%
                       {\normalfont\normalsize\itshape}}
\makeatother

\acrodef{L2R}{learning-to-rank}
\acrodef{SERP}{search result page}
\acrodef{nDCG}{normalized discounted cumulative gain}
\acrodef{PDS}{policy distributional shift}
\acrodef{OPE}{off-policy evaluation}
\acrodef{RL}{reinforcement learning}
\acrodef{RS}{recommender system}
\acrodef{SARDINE}{\underline{s}imulator for \underline{a}utomated \underline{r}ecommendation in \underline{d}ynamic and \underline{in}teractive \underline{e}nvironments}
\acrodef{SAC}{soft actor-critic}
\acrodef{VAE}{variational autoencoder}
\acrodef{HAC}{hyper-actor critic}
\acrodef{MDP}{Markov decision process}
\acrodef{POMDP}{partially observable Markov decision process}
\acrodef{PBM}{position-based model}
\acrodef{dCTR}{document click-through rate model}

\copyrightyear{2024}
\acmYear{2024}
\setcopyright{rightsretained}


\author{Romain Deffayet}
\orcid{0000-0001-8265-9092}
\affiliation{%
\institution{NAVER LABS Europe}
\city{Meylan}
\country{France}}
\affiliation{%
\institution{University of Amsterdam}
\city{Amsterdam}
\country{The Netherlands}}
\email{romain.deffayet@naverlabs.com}

\author{Thibaut Thonet}
\orcid{0000-0003-0302-0376}
\affiliation{%
\institution{NAVER LABS Europe}
\city{Meylan}
\country{France}}
\email{thibaut.thonet@naverlabs.com}

\author{Dongyoon Hwang}
\orcid{0009-0003-6471-5208}
\affiliation{%
\institution{Korea Advanced Institute of Science and Technology}
\city{Seoul}
\country{South Korea}}
\email{godnpeter@kaist.ac.kr}

\author{Vassilissa Lehoux}
\orcid{0000-0002-6512-283X}
\affiliation{%
\institution{NAVER LABS Europe}
\city{Meylan}
\country{France}}
\email{vassilissa.lehoux@naverlabs.com}

\author{Jean-Michel Renders}
\orcid{0000-0002-7516-3707}
\affiliation{%
\institution{NAVER LABS Europe}
\city{Meylan}
\country{France}}
\email{jean-michel.renders@naverlabs.com}

\author{Maarten de Rijke}
\orcid{0000-0002-1086-0202}
\affiliation{%
\institution{University of Amsterdam}
\city{Amsterdam}
\country{The Netherlands}}
\email{m.derijke@uva.nl}

\title{SARDINE: A Simulator for Automated Recommendation in Dynamic and Interactive Environments}

\begin{abstract}
Simulators can provide valuable insights for researchers and practitioners who wish to improve recommender systems, because they allow one to easily tweak the experimental setup in which recommender systems operate, and as a result lower the cost of identifying general trends and uncovering novel findings about the candidate methods. A key requirement to enable this accelerated improvement cycle is that the simulator is able to span the various sources of complexity that can be found in the real recommendation environment that it simulates. 

With the emergence of interactive and data-driven methods -- e.g., reinforcement learning or online and counterfactual learning-to-rank -- that aim to achieve user-related goals beyond the traditional accuracy-centric objectives, adequate simulators are needed. In particular, such simulators must model the various mechanisms that render the recommendation environment dynamic and interactive, e.g., the effect of recommendations on the user or the effect of biased data on subsequent iterations of the recommender system. We therefore propose \acs{SARDINE}, a flexible and interpretable recommendation simulator that can help accelerate research in interactive and data-driven recommender systems. We demonstrate its usefulness by studying existing methods within nine diverse environments derived from SARDINE, and even uncover novel insights about them.
\end{abstract}

\begin{document}		

\maketitle

\acresetall

\section{Introduction}
\label{intro}

Recommender systems must match users and items based on item content and user preferences, so as to provide users with content that fulfills a consumption need or carries relevant information given user preferences~\citep{recsys-def}. In other words, they need to learn the semantic information~\citep{semantic-information} that explains why a certain user is attracted to a certain item, usually by leveraging user features, item content or logged interactions. However, by restricting the scope of recommender systems to a static semantic matching task one would ignore a crucial part of the recommendation task: converting semantic understanding of users and items into increased value for the user, as well as for content providers and other potential stakeholders. Such value may be measured by, e.g., click-through rate, user satisfaction, retention rate, or fairness metrics.

This concern has led to the emergence of methods that consider beyond-accuracy goals~\citep{beyond-accuracy, beyond-matrix-completion} and that often view recommendation as a dynamic and interactive task~\citep{interactive-recommendation}. First, recommender systems are often trained from user interaction data, either in an online~\citep{online-recsys} or offline~\citep{unbiased-tutorial} fashion. As a result, recommender systems must learn to deal with noisy user feedback~\citep{noisy-feedback}, limited knowledge about new users in the cold start scenario~\citep{norman-beta}, as well as potential biases in user behavior that may impact the training data~\citep{unbiased-tutorial}. Second, the items consumed by a user may have an effect on the user state~\citep{spotify-diversity, gems, matthew}. They could alter user preferences -- by developing a user's interest in a topic, by educating users about a topic in a way that encourages them to explore more advanced content, or by changing their perspective on other items, for instance by sparking their interest or instead by reducing it. Items could also temporarily affect user behaviors, for instance by causing boredom, which subsequently reduces user interest and engagement in the platform~\citep{spotify-diversity}. Third, exogenous factors may change the value of items and the preferences of users, yielding an ever-changing dynamic environment~\citep{non-stationary}.

\subsection{The role of simulators in recommender systems research}

In order to account for the dynamic and interactive aspects of recommendation, various approaches have been proposed, including contextual bandits~\citep{linTS, linUCB}, \acf{RL}~\citep{topk, gems, matthew}, active learning~\citep{active-learning}, counterfactual \acl{L2R}~\citep{ips-joachims, unbiased-tutorial}, and click modeling~\citep{pbm, click-models-book, cmip}. These approaches are trained from user data, and it has been shown that they should not be evaluated solely on accuracy-centric benchmarks~\citep{offline-eval-RL4REC, beyond-matrix-completion, fresh-look-RS} as these miss the potential benefits brought by beyond-accuracy methods. 

While online evaluation~\citep{recsys-eval-survey, online-eval-recsys} remains a gold standard -- when done right~\citep{olivier-sutva} -- to evaluate the impact of recommendation models on user-related metrics, most researchers do not have access to a live recommendation system. Moreover, the potential degradation in user satisfaction and revenue induced by online experiments may limit the possibility to conduct such an evaluation, especially in a research setting where many experiments are needed to improve on the current version of the recommender system.

In that case, prior work~\citep{offline-eval-RL4REC, offpolicy-eval-slates, scope-rl} has advised to either resort to \acf{OPE}, which consists in evaluating the target system using data collected with the original system, or otherwise to conduct experiments in a simulated environment. Simulators are by definition synthetic, at least partially, and good performance obtained in a simulator is therefore no guarantee of success in the live system. However, their value lies in the ability to control relevant parameters in a way that spans the potential dynamics encountered in the real environment. Indeed, tweaking parameters and observing their effect on candidate methods allows one to identify general trends and study important research topics: regimes of success and failure (e.g., low data, high bias), robustness to environment features that may be observed in the real world (e.g., noise, distribution shifts), generalizability of the results, etc.

In that sense, simulated evaluation can even be less opaque than \acs{OPE} and online evaluation, as observing variables that are normally not accessible to the practitioner can help better interpret the observed performance of the candidate systems. In order to deliver these benefits, we argue that simulators should be:
\begin{enumerate}
    \item Configurable in a way that is easily interpretable to the practitioner,
    \item Able to span a large part of the various forms of complex behavior commonly found in the real environment.
\end{enumerate}

\noindent%
In practice, we draw up a list of specifications that we use as a goalpost for designing our simulator:

\smallskip

\noindent\textbf{Specifications \ref{specifications} --} Our simulator should satisfy the following requirements:
\label{specifications}
\begin{itemize}
    \item \textbf{Comprehensiveness:} Most of the important research questions for interactive recommender systems can be studied in one core simulated engine;
    \item \textbf{Interpretability:} One or a few well-defined parameters can control a specific aspect of interest in recommender system research, i.e., the simulator should be interpretable and controllable;
    \item \textbf{Effect isolation:} The effect of individual parameters or individual algorithmic modules can be singled out, so as to allow the focused study of one aspect of the environment (e.g., noise, user drift, etc.) or one part of the method (e.g., user and item representation, decision-making module, etc.);
    \item \textbf{Non-triviality:} The simulated task should not be trivially solved by off-the-shelf baselines; and
    \item \textbf{Configurability:} Additions and changes to the existing simulator should be easy enough to enable deeper studies or new research questions.
\end{itemize}

\noindent In order to fulfill the specifications, and before engaging with simulator design, we must define the scope of the research we wish to enable with such a simulator. We therefore define the research agenda our simulator addresses in the next section.

\subsection{A research agenda for interactive recommender systems}
\label{sec:research-agenda} 

We identify four overarching research topics (RTs) that we believe to be crucial for interactive \acfp{RS} research, and that can be studied in our simulator. We also connect them to variants of our simulator that are particularly well-suited to study them:

\begin{enumerate}[label=\textbf{(RT\arabic*)}]
    \item \textbf{How to enable multi-step reasoning and control user-related metrics in the long run?} In a dynamic and interactive environment, shifting dynamics and delayed consequences of actions prompt \acs{RS} designers to adopt a control paradigm, where target variables such as user satisfaction, revenue, or fairness-related variables must be optimized and kept at a desired value in the long run. This requires multi-step reasoning, i.e., thinking ahead of time about future consequences of recommendations formulated at the present time. Many approaches have been proposed to tackle multi-step reasoning, notably with \acl{RL}~\citep{gems, matthew, topk, prl}.This research topic can be studied thanks to the interactive environments we release, i.e., \texttt{SingleItem-Bored}, \texttt{SlateTopK-Bored}, \texttt{SlateTopK-BoredInf}, \texttt{SlateTopK}, \texttt{SlateTopK-Uncertain}, \texttt{SlateRerank-Bored}.
    
    \item \textbf{How to learn from biased data?} As online learning is often not possible in a large commercial platform, it is common to resort to offline or off-policy learning, by first collecting data in the live environment, and then learning from this data. However, multiple biases arise in the logged data. Due to selection bias, the distribution of items observed in the data is highly imbalanced, including many items that are never or almost never shown to certain users. Additionally, even when feedback is observed, biases in user behavior favor certain items above others, e.g., due to position bias. As a result, training models that do not account for these biases leads to the unfair promotion of already well-exposed items. Learning from data despite these biases is a very active area in information retrieval research, with techniques such as offline \acl{RL}~\citep{topk, matthew, prl}, counterfactual learning-to-rank~\citep{unbiased-tutorial, ips-joachims}, or click modeling~\citep{cmip, click-models-book}. All of our simulated environments can be used for off-policy training, but we notably study this research topic with our \texttt{SlateRerank-Static} and \texttt{SlateRerank-Bored} environments.
    
    \item \textbf{How to make sure that interactive recommender systems are robust to uncertainties of the real-world?} Recommender systems must operate under large amounts of uncertainty coming from multiple sources: in the user feedback and in their evolution after consuming items (e.g., varying mood and personal traits, light scanning of the results), about exogenous factors influencing user behavior and item value (e.g., world events, current context when accessing the platform), about user preferences (e.g., cold start, changing users) and in the policy itself (e.g., business rules, stochastic amortization). Large amounts of uncertainty may hurt the performance of recommender systems and yield disappointing results during the deployment of these models, which has prompted the development of uncertainty-aware methods~\citep{norman-beta, genspec}. Our \texttt{SlateTopK-Uncertain}, \texttt{SlateTopK-PartialObs} and \texttt{SingleItem-PartialObs} allow to study such uncertainties.
    
    \item \textbf{How to effectively and efficiently recommend slates (e.g., lists or grids) of items to users?} The interface of many recommendation platforms requires showing multiple recommendations to users on the same page. This comes with additional challenges as different combinations of items may lead to different short and long-term outcomes. The problem thus becomes combinatorial in nature, which makes the task intractable for most applications. The existing literature discusses slate-specific methods for both training and evaluation of slate recommendation policies~\citep{slateQ, topk, offpolicy-eval-slates}, including methods that improve on the efficiency of slate recommender systems~\citep{pl-rank, lgp}. It is possible to train slate recommender systems on all our \texttt{SlateTopK} and \texttt{SlateRerank} environments.
\end{enumerate}

\subsection{Our contributions}

Our contributions can be summarized as follows:
\begin{itemize}
    \item We introduce a \acf{SARDINE}, which can be used as a flexible core engine for multiple types of simulated experiments in recommender systems research, allowing quicker iterations towards studying, among others, the research topics (RT1--4) mentioned  in Section~\ref{sec:research-agenda}, i.e., multi-step reasoning, biased data, uncertain dynamics, and slate recommendation.
    \item We additionally provide nine different environments derived from this simulator, in the form of gymnasium~\citep{gymnasium} environments, that are already tailored for studying important aspects of recommendation in dynamic and interactive settings.\footnote{The core simulator as well as the proposed environments can be found at \href{https://github.com/naver/sardine}{https://github.com/naver/sardine}.}
    \item We conduct experiments on the nine proposed environments, in order to \begin{enumerate*}[label=(\roman*)]
        \item better describe the main dynamics of the simulator,
        \item provide a testbed for some existing approaches and baselines, and 
        \item uncover some novel findings about existing approaches, thereby restating the value of our simulator for effective recommender system research.
    \end{enumerate*}\footnote{Our experiments are open-source and can be found at \href{https://github.com/RomDeffayet/SARDINE\_Experiments}{https://github.com/RomDeffayet/SARDINE\_Experiments}.}
\end{itemize}

\noindent Furthermore, we now summarize the expected benefits of releasing our simulator. Indeed, we seek to help accelerate future research, by:
\begin{enumerate*}[label=(\roman*)]
    \item providing a playground for researchers to create and test prototypes and therefore iterate more quickly;
    \item enabling quickly building experimental set-ups in order to gain knowledge on specific research questions related to the topics RT1--4 we describe in the paper; and
    \item providing a set of not-yet-solved simulated tasks that trace a path towards progress in recommender systems research (e.g., as Atari games or Go have been for multi-step visual control).
\end{enumerate*}

In contrast, we have no intention to:
\begin{enumerate*}[label=(\roman*)]
    \item create a realistic simulator of the human mind -- besides clearly being an unattainable goal, we argue that it is not necessary to gain perfect knowledge of the actual underlying user model to effectively optimize the target variables (e.g., user engagement). Instead, we propose to study the adaptability and robustness of recommendation agents, with the help of a large array of different simulated settings.
    \item Provide guarantees of live performance. Simulators, whether they are fully- or semi-synthetic, cannot provide guarantees of performance in the live recommender system. They are nonetheless valuable for making progress in recommender systems research, e.g., by studying the robustness of agents and the edge cases where they might struggle, by quickly iterating on simulated tasks that robust recommenders should be able to solve, or even by detecting poorly robust methods before conducting A/B testing in a live system and potentially negatively impacting real users. And
    \item replace offline evaluation on traditional metrics. While a set of diverse simulated experiments offers a unique perspective on the inner workings of recommender systems, simulations must always be complemented with offline and online real-world experiments in order to build a well-rounded assessment of the progress in recommender systems research.
\end{enumerate*}

The remainder of the paper is organized as follows. We formally define the recommendation problem of interest in Section~\ref{problem-definition}. We then describe the technical details of the SARDINE simulator in Section~\ref{simulator}. Section~\ref{experimental-setup} covers the details about our experimental setup, which includes the description of the SARDINE environments tested in our experiments as well as the compared approaches. The experiment results are presented and discussed in Section~\ref{experiments}. Finally, we compare our proposed SARDINE to existing recommendation simulators in Section~\ref{related-work}, and conclude the paper in Section~\ref{conclusion}.

\section{Problem definition}
\label{problem-definition}

The problem studied in this paper can be defined as slate recommendation\footnote{We consider that single-item recommendation is just a special case of slate recommendation with a slate of size one. Therefore our problem formulation also covers this case.} in a dynamic environment. In this scenario, we consider that a user interacts with a recommender system over a session of $L$ steps. In each step, the recommender system presents a slate containing $S$ items from a predefined set $\mathcal{I}$ of cardinal $n_{\mathcal{I}}$ to the user. Based on the affinity between the recommended items and the user preferences, the user decides to click on some or none of the slate items. Information about the interaction and the current user state is then returned to the agent and, based on this, the recommender determines the next slate to recommend. This process can be formulated as a \acf{MDP} $\mathcal{M} = (\mathcal{S}, \mathcal{A}, P, R)$ defined as follows:
\begin{itemize}
    \item A set of states $s \in \mathcal{S}$, which represent the user state and summarize information about the past interactions.
    \item A set of actions $a \in \mathcal{A}$ corresponding to the possible slates presented by the recommender to the user. This set covers all slates combining items from $\mathcal{I}$, so that $|\mathcal{A}| = \frac{n_{\mathcal{I}}!}{(n_{\mathcal{I}} - S)!}$ for a slate of size $S$.
    \item A set of transition probabilities $P : \mathcal{S} \times \mathcal{A} \times \mathcal{S} \rightarrow [0,1]$, which define the dynamics in the process, i.e., how likely a state $s' \in \mathcal{S}$ is if the recommender takes action $a \in \mathcal{A}$ in state $s \in \mathcal{S}$.
    \item A (potentially stochastic) reward function $R : \mathcal{S} \times \mathcal{A} \rightarrow \mathbb{R}$, which we define as the sum of clicks over the recommended slate.
\end{itemize}

\noindent We also define a possibly stochastic policy $\pi : \mathcal{S \times A} \rightarrow [0,1]$ whose role is to decide what slate $a$ the recommender system should return in a given state $s$. A trajectory $\tau$ is defined as the set of successive states, actions and rewards collected in a session of interactions between a user and a recommender. We denote as $\tau \sim \pi$ the fact that trajectory $\tau$ is generated by following the actions provided by policy $\pi$. The problem of slate recommendation in a dynamic environment can then be summarized as identifying a policy $\pi^*$ that maximizes the cumulated reward (also known as \textit{return}) in expectation over possible trajectories, i.e., $\pi^* \in \mathrm{arg}\max_\pi \mathbb{E}_{\tau \sim \pi} \left[ \sum_{(s, a) \in \tau} R(s, a) \right]$.

In this paper, we introduce a simulator that instantiates the MDP described above to provide a testbed for developping recommendation policies and studying their characteristics in various settings. The proposed simulator is further described in Section~\ref{simulator}.

\section{Simulator}
\label{simulator}

\begin{figure}[t!]
\centering
\includegraphics[width=0.95\textwidth]{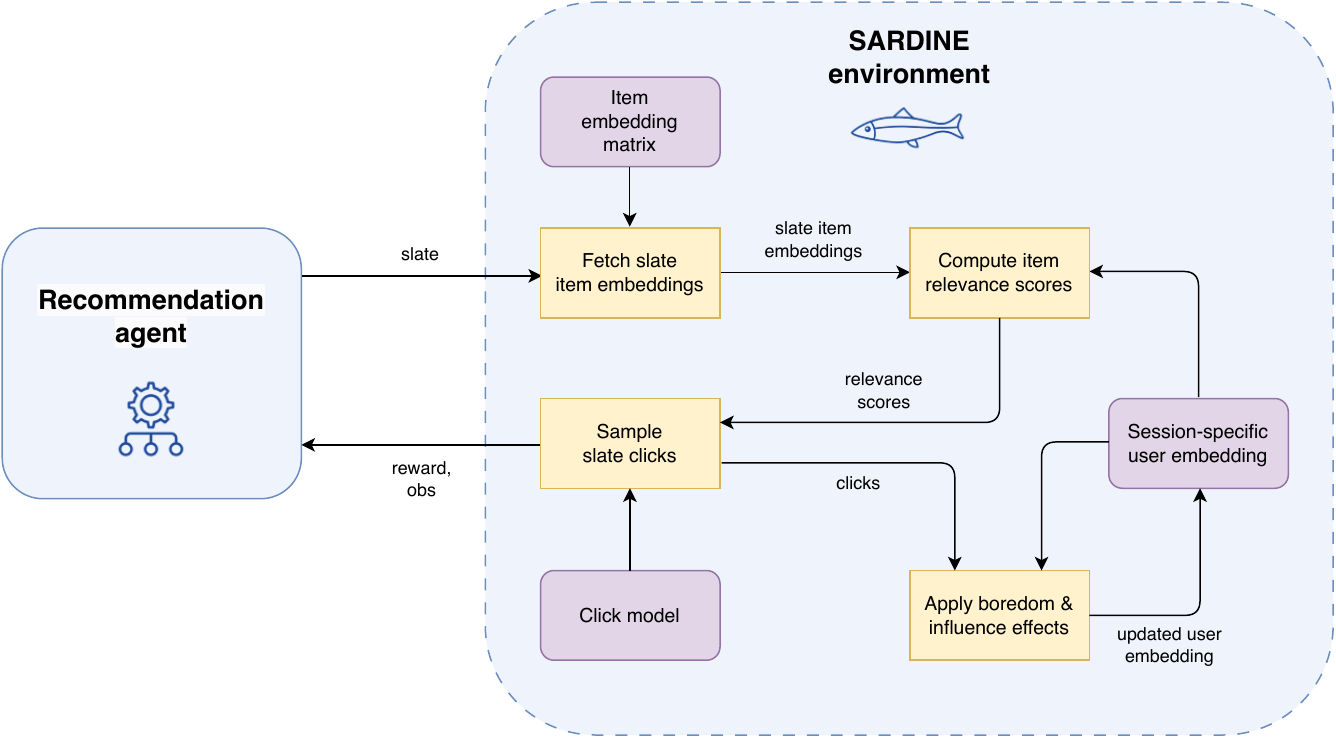}
\vspace{0.2cm}
\caption{Diagram summarizing the different components of the proposed SARDINE simulator, and its interaction with the recommendation agent.}
\label{fig:slatesim}
\end{figure}

In this section, we detail the components of our proposed \underline{s}imulator for \underline{a}utomated \underline{r}ecommendation in \underline{d}ynamic and \underline{in}teractive \underline{e}nvironments, or SARDINE in short. In SARDINE, we consider a cold-start scenario where each new session corresponds to a new user, generated on-the-fly. This means that we assume no prior knowledge on user profiles before a session starts and that the agent must do some exploration to discover user interests. This design choice is realistic for many recommendation platforms, e.g., when a single device or profile regroups several users~-- who exhibit diverse preferences over different sessions~-- or when the platform does not track a user ID for privacy reasons \cite{Hidasi2016}.

First, our simulator is initialized by forming synthetic embeddings for the set of recommendable items (Section \ref{sec:embeddings}). Then, each user session is generated by following these successive steps:
\begin{enumerate}
    \item Sample a user embedding for the current session's user (Section \ref{sec:embeddings});
    \item Provide an initial recommendation (Section \ref{sec:init}) or prompt the agent to recommend a slate to the user;
    \item Compute the relevance of the items in the slate with respect to the user and sample the clicks on the slate based on items' relevance and rank (Section \ref{sec:click});
    \item Update the user embedding to account for the effects of boredom and clicked item influence, if those mechanisms are included in the simulator (Section \ref{sec:boredom});
    \item Repeat steps (2) to (4) until the number of interaction steps reaches the session length $L$.
\end{enumerate}
We define both a fully observable variant and a partially observable variant for SARDINE, whose differences are detailed in Section \ref{sec:observability}. Moreover, we use the main engine described in this section with different sets of hyperparameters so as to create nine different environments with various characteristics, and targeting various research outcomes. We introduce these environments in Section~\ref{sec:env}.

Fig.~\ref{fig:slatesim} illustrates the different components of our simulator and its interactions with the recommendation agent. In Table~\ref{tab:hyperparam-desc} we additionally provide a description for the hyperparameters of the simulator, which are further defined in the remainder of this section.

\begin{table}[t!]
\centering
\caption{List of the hyperparameters used in the proposed SARDINE simulator, with their description.}
\vspace{0.2cm}
\label{tab:hyperparam-desc}
\begin{tabular}{@{}cl@{}}
\toprule
Hyperparameter & Description \\ \midrule
$L$ & Session length (in time steps). \\
$S$ & Slate size (in number of items). \\
$n_{\mathcal{I}}$ & Number of items. \\
$n_\mathcal{T}$ & Number of topics (and user/item embedding dimension). \\
$\lambda$ & Scale hyperparameter for the relevance function. \\
$\mu$ & Shift hyperparameter for the relevance function. \\
$\alpha$ & Range hyperparameter for item attractiveness. \\
$\epsilon$ & Click propensity for examination probability. \\
$n_\text{b}$ & Number of items considered for boredom computation. \\
$t_\text{b}$ & Click recency (in time steps) for boredom computation. \\
$\tau_\text{b}$ & Threshold on topic occurrence for boredom computation. \\
$\omega$ & Weight controlling the influence of clicked items on user. \\
$\mathcal{O}$ & Hyperparameter indicating full or partial state observability. \\ \bottomrule
\end{tabular}
\end{table}

\subsection{Item and user embeddings}
\label{sec:embeddings}

Items and users are assigned randomly-generated sparse embeddings of size $n_{\mathcal{T}} = |\mathcal{T}|$, where $\mathcal{T}$ is the set of topics associated to items and users (defined below). The sparsity enforces a coverage of only a limited number of topics per item and user. The generative process to define the embedding\footnote{To distinguish the item embeddings used in the simulator from the item embeddings that may be learned by an agent, we refer to the former as \textit{ideal item embeddings} when disambiguation is needed.} for each item $i$ in the set of items $\mathcal{I}$ is the following:
\begin{enumerate}
    \item Sample the item embedding components from a uniform distribution over $[0, 1]$: $\mathbf{e}_i = (\mathbf{e}_{i,1}, \ldots, \mathbf{e}_{i,n_{\mathcal{T}}}) \in \mathbb{R}^{n_{\mathcal{T}}}$ with $\mathbf{e}_{i,j} \sim \text{Unif}([0, 1])$;
    \item Sample a number of topics associated to the item (i.e., the number of non-zero components to retain in $\mathbf{e}_i$) equal to either 2 or 3: $n_{\mathcal{T}_i} \sim \text{Unif}(\{2, 3\})$;
    \item Sample the $n_{\mathcal{T}_i}$ topics associated to the item from the topic set $\mathcal{T}$: $\mathcal{T}_i = \{T_{i,1}, \ldots, T_{i,n_{\mathcal{T}_i}}\} \subset \mathcal{T}$ with $T_{i,j} \sim \text{Unif}(\mathcal{T})$ without replacement;
    \item Zero out the components of the item embedding that correspond to non-selected topics (i.e., outside of $\mathcal{T}_i$): $\mathbf{e}_{i,j} := 0$ if $j \in \mathcal{T} \setminus \mathcal{T}_i$;
    \item Normalize the components to have an embedding with unitary Euclidean norm: $\mathbf{e}_i := \frac{\mathbf{e}_i}{\lVert\mathbf{e}_i\rVert_2}$.
\end{enumerate}
We denote the \emph{main topic} of item $i$ as $T^*_i$ which corresponds to the dominating component in the item embedding $\mathbf{e}_i$, i.e., $T^*_i = \mathrm{arg}\max_{j \in \mathcal{T}} \mathbf{e}_{i,j}$. 

The process to generate a user embedding $\mathbf{e}_u$ for each new session is similar to that of generating an item embedding, with the difference that we allow a user embedding to cover 3, 4, or 5 topics instead of 2 or 3 for items. The rationale for this choice is that a user may be interested in a broad selection of topics whereas an item usually spans a more narrow set of topics (e.g., the number of movie genres a user likes vs.\ the number of genres a movie belongs to). 

\subsection{Initial recommendation}
\label{sec:init}

The first recommendation the user receives at the beginning of a session is independent of the agent and done directly in the simulator. Given our cold-start setting, i.e., we have no prior knowledge of the user profile, we wish to start the session by probing the preferences of the user. For that purpose, our initial recommendation is simply a slate containing random items. In a real-life scenario, other alternatives could be considered, e.g., by exploiting the popularity of items~\citep{warm-start-popularity}, prior user profile~\citep{twitter-warm} as well as user metadata~\citep{warm-start-autoencoder}. We leave the investigation of alternative initial recommendations for future work.

\subsection{Relevance computation \& click model}
\label{sec:click}

In this section, we describe how relevance, i.e., the matching score, is computed for a (user, item) pair. Then we detail how this relevance score is used to sample the clicks and skips on a slate recommended to the user.

\paragraph{\bf Relevance score.}

The relevance of items presented to a user is calculated based on the dot-product between the item embedding and the user embedding:
\begin{equation}
    \mathrm{rel}(i, u) = \mathbf{e}_i^T \mathbf{e}_u.
\end{equation}

\paragraph{\bf Item attractiveness.}

To the relevance score we then apply a sigmoid function that is rescaled and shifted to account for the range of values and the desired level of saturation for the function, resulting in an attractiveness score. Compared to the relevance score, the attractiveness of an item reflects click behavior specified by the hyperparameters of the sigmoid; their role is explained below. 
Formally, the attractiveness of item $i$ for user $u$ is defined as follows: 
\begin{equation}
    \label{eq:attractiveness}
    A_{u,i} = \alpha \cdot \sigma(\mathrm{rel}(i, u)) \, \text{ where } \, \sigma(x) = \frac{1}{1 + \exp(-\lambda (x-\mu))}.
\end{equation}
The hyperparameter~$\alpha$ is introduced to adjust the range of the attractiveness score. The shift hyperparameter $\mu$ ensures that the function outputs a value close to 1 for a highly matching (user, item) pair, and close to 0 for an item totally unrelated to the user. The scale hyperparameter $\lambda$ controls how steep the sigmoid will be (i.e., how easily the output of the function saturates to 0 or 1). This latter hyperparameter plays a key role for the level of uncertainty in the simulator. Indeed, a lower value of $\lambda$ implies that the sigmoid $\sigma$ will be less steep, leading to smaller differences in attractiveness (and, in turn, in click probabilities as detailed below) between relevant and irrelevant items. In other words, the user feedback is more uncertain when $\lambda$ is low.

In practice, we set the values of $\lambda$ and $\mu$ using the following rules: 
\begin{enumerate}
    \item A random recommendation policy should almost always propose irrelevant items, i.e., such that $\sigma({\mathbf{e}_i}^T \mathbf{e}_u )$ is generally close to 0 for a randomly selected item $i$;
    \item An oracle recommendation policy should always propose relevant items, i.e., such that $\sigma({\mathbf{e}_i}^T \mathbf{e}_u )$ is close to 1 for the few top items $i$ that best match user $u$;
    \item A bored\footnote{The notion of boredom introduced in this simulator is further detailed in Section~\ref{sec:boredom}.} user cannot be satisfied most of the time, even by an oracle recommendation policy, i.e., when user $u$ is in a bored state, $\sigma({\mathbf{e}_i}^T \mathbf{e}_u )$ is much smaller than 1 even for the few top items $i$ that best match $u$.
\end{enumerate}

\paragraph{\bf Click model.}

After the attractiveness score for a (user, recommended item) pair has been computed for each item of the slate, the simulator has to decide if this pair leads to a click or not. For that purpose, we consider a position-based click model, i.e., the probability of click is defined by the product of item-specific attractiveness and rank-specific examination probability. More complex click models could be considered and added to the simulator, but we do not wish to provide a catalog of all existing models. Instead, we want to show that the impact of biased data in general is visible in our simulator, taking the position-based model as an example.

Formally, this click probability is expressed as $\mathbb{P}(c \mid u, i, r) = A_{u,i} \times E_r$ for an item $i$ positioned at rank $r \in \{1, \ldots, S\}$ in the slate, with $S$ the slate size. $A_{u,i}$ is the attractiveness of item $i$ for user $u$ defined in Eq.~\ref{eq:attractiveness} and $E_r$ is the probability that the user examines the items in the slate down to rank $r$. By default, the examination probability is set to $E_r = \varepsilon^{r-1}$, where the hyperparameter $\varepsilon$ defines the rate of decay of the examination probability.  The click (or skip) from user $u$ on item $i$ at rank~$r$ in the slate is then sampled from the Bernoulli distribution $\text{Bern}(A_{u,i} \times E_r)$.

\subsection{Boredom and influence mechanisms}
\label{sec:boredom}

Our SARDINE simulator introduces two long-term mechanisms in the recommendation, which penalize myopic strategies and thus require the agent to consider the consequences of its actions several steps after taking them. The rationale for this choice is to be able to generate benchmarks where reinforcement learning-based agents are a better choice than bandit approaches, which would otherwise be more suitable for a greedy sequential recommendation task as shown in \cite{Lee2022}. This goal is motivated by empirical evidence of the limits of greedy methods with respect to, e.g., diversity, and thus their detrimental impact on long-term metrics such as churn rate~\citep{matthew, spotify-diversity, spotify-balancing}. The first mechanism we define is referred to as \textit{boredom} and intuitively reflects the fact that a user may become less interested in consuming content (i.e., clicking on items) when the items recommended in successive slates are too similar, similarly to \citep{matthew, gems}. The second mechanism we consider is the \textit{influence of the clicked items} on the future user behavior: when a user consumes an item, this may shift the user's interest towards the item's topics, as in~\citep[e.g.,][]{echo-chambers}. These two mechanisms are described in more detail below.

\paragraph{\bf Boredom.} To determine if a user $u$ gets bored during a session, we consider the items clicked in the last $t_\text{b}$ time steps. 
If there are more than~$n_\text{b}$ such items, we keep only the $n_\text{b}$ most recently clicked items and we record a list of their main topics.
Then, if a topic $T \in \mathcal{T}$ occurs more than a threshold of $\tau_\text{b} \leq n_\text{b}$ times in this list, we consider that the user $u$ is bored with respect to topic~$T$. We define two boredom variants that specify the impact on the bored user's behavior: \textit{temporary loss-of-interest boredom} and \textit{churn-and-return boredom}.
For the temporary loss-of-interest boredom, the user $u$ who is bored with respect to topic $T$ has their user embedding component $\mathbf{e}_{u,T}$ set to 0 (i.e., this simulates a loss of interest for topic $T$) for $t_\text{b}$ time steps. After this period, we consider that the boredom effect has timed out and the user may be willing to click again on items with main topic $T$, so the component $\mathbf{e}_{u,T}$ is restored to its previous value. The churn-and-return boredom operates in a similar fashion with the difference that all components of $\mathbf{e}_u$ are set to 0 until the boredom effect times out: this simulates the fact that the user churns the platform (as an all-zero user embedding implies an absence of clicks in our simulator) and then returns after $t_\text{b}$ time steps.

\paragraph{\bf Clicked item influence.}

At each interaction step, the user $u$ is recommended a slate and potentially clicks on some of this slate's items. We denote as $\mathcal{I}_c$ this set of clicked items. We transcribe the influence of clicked items on $u$'s future behavior by updating the user embedding $\mathbf{e}_u$ as a weighted average of the previous user embedding and the mean of the clicked item embeddings: $\mathbf{e}_u := \omega \, \mathbf{e}_u + (1 - \omega) \frac{1}{|\mathcal{I}_c|}\sum_{i \in \mathcal{I}_c}\mathbf{e}_i$, where $\omega$ is a hyperparameter controlling the amount of influence clicked items have on the user. Intuitively, the influence mechanism causes a drift in user interests and thus makes the recommendation process more dynamic.

\subsection{Full observability vs partial observability}
\label{sec:observability}

Our SARDINE simulator can be used in two modes: either with \textit{full} state observability or with \textit{partial} state observability, which is recorded in the hyperparameter $\mathcal{O}$. The former simulates a \acf{MDP} setting while the latter defines a \acf{POMDP} setting. In this section, we define the state/observation used in these two cases.

\paragraph{\bf Full observability.} In the full observability case, agents have access to the entire information about the user state. Here, the state fed to the agent is defined as the concatenation of 3 vectors:
\begin{itemize}
    \item The current user embedding, i.e., $\mathbf{e}_u$, which corresponds to the user embedding at the current time step and thus includes the effects of boredom and influence (if those mechanisms are included in the simulator). Size: $n_{\mathcal{T}}$.
    \item A histogram indicating the number of times each topic was the main topic of an item among the $n_b$ last clicked items in the most recent $t_b$ time steps. The histogram is normalized by dividing click numbers by the threshold $\tau_b$ and by clipping between 0 and 1. Size: $n_{\mathcal{T}}$.
    \item A vector indicating the boredom timeout duration (in number of steps) left for each topic. If a topic is not in a bored state for the user, then its default timeout duration is $t_b$. For topics that triggered boredom in previous steps and whose boredom is still ongoing, the duration will be between 0 and $t_b$ (excluded). This vector is also normalized between 0 and 1 by dividing it by $t_b$. Size: $n_{\mathcal{T}}$.
\end{itemize}
In the state, the current user embedding is used to keep track of the dynamic user preferences, while the histogram and timeout vectors maintain the information about recent item consumption and boredom. The current item embeddings~-- which represent the actual preferences of a user at a given time~-- are normally not available in a real-life recommendation scenario. However, studying this fully observable setting enables the practitioner to single out the impact of the recommendation algorithm, contrarily to the partially observable setting which compounds the effects of algorithm effectiveness and user embedding estimation quality. As we will show in our experiments (Sections~\ref{experimental-setup} and~\ref{experiments}), the fully observable case already leads to challenging environments which justifies our choice to include this less realistic scenario.

\paragraph{\bf Partial observability.} For the partial observability setting, the agent cannot access the inner workings of the simulator and is only provided a set of observations about the last interaction. The observation returned to the agent is the concatenation of 3 vectors:
\begin{itemize}
    \item The slate that was recommended by the agent, with the item ID for each slot. Size: $S$.
    \item The clicks that the user did on the recommended slate, with 1 or 0 at each slot to indicate a click or a skip, respectively. Size: $S$.
    \item The histogram of recent clicked topics, as in the fully observable case. It is realistic to consider this information accessible to the agent as item categories in recommender systems are generally public. Size: $n_{\mathcal{T}}$.
\end{itemize}
Based on these 3 pieces of information, the agent is able to identify which recommended items led to a click and exploit recently clicked topics to better infer user preferences. However, they are not enough to perfectly determine the user state and the agent may need to incorporate the history of observations in the same session in order to improve its estimation of the user state (which is usually done through state encoders).

\section{Experimental Setup}
\label{experimental-setup}

Now that we have detailed the main components of our simulator in the previous section, we can describe some of its possible instantiations for conducting experiments related to the research agenda of Section~\ref{sec:research-agenda}. This section therefore aims
\begin{enumerate*}[label=(\roman*)]
    \item to provide guidance for the usage of the simulator,
    \item to define a testbed for studying existing methods along the research topics defined in Section~\ref{sec:research-agenda}, and
    \item to demonstrate the simulator's utility for recommendation research by uncovering some novel insights about these methods.
\end{enumerate*} 

First, Section~\ref{sec:env} introduces nine recommendation environments instantiated from our simulator, that we use in our experiments. Then, Section~\ref{sec:baselines} describes the recommendation agents we seek to compare on the environments. Finally, Section~\ref{sec:hyperparam} summarizes the simulator hyperparameters adopted by the different environments, as well as the agent hyperparameters used in our experiments.

\subsection{Simulated environments}
\label{sec:env}

To demonstrate possible use cases enabled by SARDINE, we defined nine different environments~-- each being a variant of our simulator. The characteristics of these nine environments are detailed in Table~\ref{tab:env}. 
They are characterized along six dimensions, which are directly linked to the research topics defined in Section~\ref{sec:research-agenda}:
\begin{enumerate}[label=(\arabic*)]
    \item The type of recommendation made to the user: single-item recommendation (corresponding to the case where $S = 1$) or slate recommendation ($S > 1$)~--- \textbf{RT4};
    \item The presence of a boredom mechanism, i.e., users get bored when being presented repeatedly with a similar content, and thus become less likely to click on the related items~--- \textbf{RT1};
    \item The presence of an influence mechanism, i.e., users are influenced by clicked items in future interaction steps~--- \textbf{RT1};
    \item The level of click uncertainty, i.e., the degree of stochasticity in the click probabilities, which is controlled by the scale hyperparameter $\lambda$ in the relevance sigmoid: lowering $\lambda$ increases the click likelihood on less relevant items (see Section~\ref{sec:click} for more details)~--- \textbf{RT3};
    \item The observability, i.e., whether the agent has access to full or partial user state information (MDP or POMDP setting, respectively) as detailed in Section~\ref{sec:observability}~--- \textbf{RT3};
    \item Whether the task is reranking, in which case there is a limited number of items that are all presented to the user (i.e., $n_{\mathcal{I}} = S$) and the recommendation agent has to find the best permutation of those items\footnote{An example of such a scenario in a real-life recommender system is in a two-stage setting where the recommender first reranks the (limited) set of item categories for the user. Then, in a second step, the recommender identifies the best item to present for each ranked category slot. What we are interested in here is the first reranking step where items correspond to categories.}~--- \textbf{RT2}.
\end{enumerate}
Below, we summarize the purpose of each of the nine environments we introduce:
\begin{itemize}
    \item \texttt{SingleItem-Static}: This single-item recommendation environment with static user behavior and full state observability was chosen to showcase an ``easy'' environment where learned agents should be able to reach optimal performance without difficulty. This environment also provides a good sanity check to validate that a learned agent is working as expected.
    \item \texttt{SingleItem-BoredInf}: This environment augments \texttt{SingleItem-Static} with boredom and influence long-term mechanisms, which require the agent to consider multi-turn dynamics to provide effective recommendations. Therefore, this corresponds to a typical RL-based recommendation environment, in an MDP setting.
    \item \texttt{SingleItem-PartialObs}: This is another variant of \texttt{SingleItem-Static} that increases the environment's difficulty through partial observability, i.e., the true state is not directly accessible and the agent is only provided with partial observations at each interaction step. This simulates typical sequential recommendation environments based on offline feedback, where the state (i.e., the user embedding) is unknown and recommendations have no causal effect on future user interactions \cite{offline-eval-RL4REC}.
    \item \texttt{SlateTopK-Bored}: This variant of the simulator includes slate recommendation (as opposed to the single-item recommendation from the previous environments) and a boredom mechanism, with full state observability. It makes this environment suitable to evaluate RL-based slate recommendation methods in an MDP setting.
    \item \texttt{SlateTopK-BoredInf}: This environment is based on \texttt{SlateTopK-Bored} with an additional influence mechanism, making the dynamics more complex as clicked items' influence causes a drift in user interests.
    \item \texttt{SlateTopK-PartialObs}: This challenging environment derived from \texttt{SlateTopK-BoredInf} includes boredom and influence mechanisms, but also partial observability. The POMDP setting along with the need for RL-based agents to tackle the effects of the long-term mechanisms make this environment a good choice to investigate state encoders as well as RL-based slate recommendation agents.
    \item \texttt{SlateTopK-Uncertain}: In this environment, we start from \texttt{SlateTopK-PartialObs} and increase the uncertainty through greater stochasticity in the clicking process. In practice, this is done by reducing the value of the relevance scale hyperparameter $\lambda$. We vary $\lambda$ from its standard value 100 (used in \texttt{SlateTopK-PartialObs}) to 10, 5 or 2 to study different levels of click uncertainty.
    \item \texttt{SlateRerank-Static}: This environment is focused on the reranking task described previously and includes static users. Its main purpose is to enable us to study the effect of the ranking order (i.e., the presentation bias) as opposed to the mere effect of including items in the ranking (i.e., the selection bias), as done in \texttt{SlateTopK} environments. This environment and its potential variants are therefore particularly suited for click modeling and counterfactual learning-to-rank research.
    \item \texttt{SlateRerank-Bored}: Similarly to \texttt{SlateRerank-Static}, this environment provides a test\-bed for research on presentation biases such as position bias. However, it adds a boredom mechanism so that greedy agents, even with perfectly alleviated position bias, are not optimal. It thus constitutes a way to conduct research on the effect of data biases on, e.g., RL agents.
\end{itemize}

\noindent%
The set of environments introduced above is not intended to give an exhaustive coverage of all possible combinations allowed by our simulator, but rather to provide a sample of relevant environments highlighting its various possibilities. In particular, we chose these environments to reflect the four research topics introduced in Section~\ref{sec:research-agenda}: the inclusion of multi-step mechanisms (in \texttt{SingleItem-BoredInf}, \texttt{SlateTopK-Bored}, \texttt{SlateTopK-BoredInf}, \texttt{SlateTopK-PartialObs}, and \texttt{SlateRerank-Bored}), the biases induced by the item presentation order (in \texttt{SlateRerank-Static} and \texttt{SlateRerank-Bored}), the uncertainty in the clicks (in \texttt{SlateTopK-Uncertain}) and in the user state (in \texttt{SingleItem-PartialObs}, \texttt{SlateTopK-PartialObs}, and \texttt{SlateTopK-Uncertain}), and the recommendation of slates as opposed to single items (\texttt{SlateTopK} and \texttt{SlateRerank} environments vs.\ \texttt{SingleItem} environments). The precise set of hyperparameters used in each environment are detailed in Section~\ref{sec:hyperparam}.

As a side note, in Section~\ref{sec:boredom}, we defined two types of boredom mechanism: the temporary loss-of-interest boredom and the churn-and-return boredom. In our experiments, we only use the churn-and-return boredom. Indeed, the experiments done in our pilot studies with the two boredom mechanisms lead to similar conclusions on the approaches' relative performance. Therefore, we omit results with the temporary loss-of-interest boredom for the sake of brevity. 

\begin{table}[t]
\centering
\caption{Description of the nine environments studied in our experiments, each corresponding to a variant of our simulator.}
\vspace{0.2cm}
\label{tab:env}
\scalebox{0.85}{
\begin{tabular}{@{}lllllll@{}}
\toprule
Environment name & Rec. type & Boredom & Influence & Click uncertainty & Observability & Reranking \\ \midrule
\texttt{SingleItem-Static} & Single item & No & No & Low & Full & No \\
\texttt{SingleItem-BoredInf} & Single item & Yes & Yes & Low & Full & No \\
\texttt{SingleItem-PartialObs} & Single item & No & No & Low & Partial & No \\
\texttt{SlateTopK-Bored} & Slate & Yes & No & Low & Full & No \\
\texttt{SlateTopK-BoredInf} & Slate & Yes & Yes & Low & Full & No \\
\texttt{SlateTopK-PartialObs} & Slate & Yes & Yes & Low & Partial & No \\
\texttt{SlateTopK-Uncertain} & Slate & Yes & Yes & Medium to v. high & Partial & No \\
\texttt{SlateRerank-Static} & Slate & No & No & High & Full & Yes \\
\texttt{SlateRerank-Bored} & Slate & Yes & No & High & Full & Yes \\ \bottomrule
\end{tabular}
}
\end{table}

\subsection{Compared methods}
\label{sec:baselines}

This section presents the different baseline recommendation methods that we re-implemented in SARDINE\footnote{The implementation for those methods is included in our code at \href{https://github.com/RomDeffayet/SARDINE\_Experiments}{https://github.com/RomDeffayet/SARDINE\_Experiments} and made available for the sake of reproducibility.} and tested in our experiments. We sought to include both simple, naive baselines as well as recent and state-of-the-art approaches to highlight the different characteristics and difficulty levels of the environment presented in Section~\ref{sec:env}. Our compared methods include the following:
\begin{itemize}
    \item \textbf{Random:} This simple baseline simply consists in recommending a random slate (or item in the case of \texttt{SingleItem} environments) at each interaction step.
    \item \textbf{Greedy Oracle:} This baseline recommends at each step the optimal slate (or item in the case of \texttt{SingleItem} environments) based on the current user embedding. The optimal slate contains the $S$ items that maximize the relevance function defined in Section~\ref{sec:click}, ordered by relevance in a top-down fashion. This approach is optimal in a static setting (without boredom and influence). However, it is unable to perform multi-step reasoning in a dynamic setting (with boredom and/or influence) due to its myopic behavior, hence the name Greedy Oracle.
    \item \textbf{REINFORCE + Top-K:} This approach proposed in \cite{topk} extends the REINFORCE policy-gradient agent to the slate recommendation problem. It estimates the value of individual items rather than the full slate, thereby making the problem tractable. However, it requires certain assumptions, for instance that the slate receives at most one click and that the items' returns are mutually independent. Since slates can have several clicks in SARDINE, we simply use the first click in the slate for this method. For the \texttt{SingleItem} environments, we instead use a standard REINFORCE agent as the top-K addition is not needed.
    \item \textbf{SAC + Top-K:} This method was introduced in \cite{gems} as a simple yet strong baseline for slate recommendation. It relies on a \acf{SAC} \cite{Haarnoja2018} policy that takes actions in the item embedding space. The recommended slate is then formed by identifying the items which maximize the dot-product with the action, i.e., the K-nearest neighbors, and by ordering them in a top-down fashion. For the \texttt{SingleItem} environments, we adopt a standard SAC agent and simply replace the top-K selection by a top-1 selection.
    \item \textbf{SAC + GeMS:} Proposed in \cite{gems}, this approach relies on a \acf{VAE} to embed the high-dimensional slate space into a low-dimensional latent space, that is used as a tractable action space for a SAC agent. This process is done in two steps. First, a VAE is trained on logged data containing past user sessions with slates and clicks. For that purpose, we generate a dataset which collects interactions between the environment of interest and a logging policy corresponding to a uniformly balanced mixture of a Random agent and a Greedy Oracle agent.\footnote{In other words, each item in a slate generated by the logging policy has 50\% chance to be the item the Greedy Oracle would recommend at this rank, and 50\% chance to be a random item.} Second, the frozen decoder of the VAE is plugged on the output of a SAC agent to reconstruct a slate from the agent's action in the latent space. 
    \item \textbf{HAC:} Similarly to the GeMS framework, the \acf{HAC} method \cite{Liu2023} proposes to use an RL agent which takes actions in a latent space and introduces a module to translate latent actions into slates. Differently from SAC + GeMS, this approach relies on the DDPG~\cite{Lillicrap2015} policy and it does not exploit a VAE to regularize the latent space. It also requires no pretraining as all parameters are learned in an off-policy fashion. Moreover, HAC uses a supervised click prediction objective in addition to the RL one, in order to stabilize the learning of the agent and directly exploit the user response signal on slate items.
\end{itemize}

\noindent%
In our experiments, we consider that methods have access to the ideal item embeddings, i.e., the item embeddings that are used in the simulator (whose generation was described in Section~\ref{sec:embeddings}). This constitutes an advantage for the agents which explicitly use item embeddings in their method, namely, Greedy Oracle, SAC + Top-K, SAC + GeMS, and HAC. The other approaches (Random and REINFORCE + Top-K) therefore have a slight disadvantage over the former methods for that reason. To study the impact of the access to high-quality item embeddings, we also compared the results with ideal embeddings to those obtained using sub-optimal, matrix factorization embeddings (see the experiments on \texttt{SlateTopK-Bored} in Section~\ref{sec:exp-topk}).


\subsection{Hyperparameter setting}
\label{sec:hyperparam}

\begin{table}[t]
\centering
\caption{Value of the simulator hyperparameters for each of the nine environments used in our experiments. The description of the hyperparameters' meaning and role is detailed in Table~\ref{tab:hyperparam-desc}. An N/A value signals that the phenomenon related to the hyperparameter is absent in this environment (e.g., the influence parameter $\omega$ is N/A for the \texttt{SlateTopK-Bored} environment which does not include the influence mechanism).}
\vspace{0.2cm}
\label{tab:hyperparam-val}
\scalebox{0.85}{
\begin{tabular}{@{}llllllllllllll@{}}
\toprule
\multirow{2}{*}{Environment name} & \multicolumn{12}{c}{Hyperparameter value} &  \\ \cmidrule(l){2-14} 
 & $L$ & $S$ & $n_{\mathcal{I}}$ & $n_{\mathcal{T}}$ & $\lambda$ & $\mu$ & $\alpha$ & $\epsilon$ & $n_{\text{b}}$ & $t_{\text{b}}$ & $\tau_{\text{b}}$ & $\omega$ & $\mathcal{O}$ \\ \midrule
\texttt{SingleItem-Static} & 100 & 1 & 1000 & 10 & 100 & 0.65 & 1.0 & 0.85 & N/A & N/A & N/A & 1.0 & full \\
\texttt{SingleItem-PartialObs} & 100 & 1 & 1000 & 10 & 100 & 0.65 & 1.0 & 0.85 & N/A & N/A & N/A & 1.0 & partial \\
\texttt{SingleItem-BoredInf} & 100 & 1 & 1000 & 10 & 100 & 0.65 & 1.0 & 0.85 & 10 & 5 & 5 & 0.95 & full \\
\texttt{SlateTopK-Bored} & 100 & 10 & 1000 & 10 & 100 & 0.65 & 1.0 & 0.85 & 10 & 5 & 5 & 1.0 & full \\
\texttt{SlateTopK-BoredInf} & 100 & 10 & 1000 & 10 & 100 & 0.65 & 1.0 & 0.85 & 10 & 5 & 5 & 0.95 & full \\
\texttt{SlateTopK-PartialObs} & 100 & 10 & 1000 & 10 & 100 & 0.65 & 1.0 & 0.85 & 10 & 5 & 5 & 0.95 & partial \\
\texttt{SlateTopK-Uncertain} & 100 & 10 & 1000 & 10 & \{2, 5, 10\} & 0.65 & 1.0 & 0.85 & 10 & 5 & 5 & 0.95 & partial \\
\texttt{SlateRerank-Static} & 10 & 10 & 10 & 10 & 5 & 0.30 & 1.0 & 0.85 & N/A & N/A & N/A & 1.0 & full \\
\texttt{SlateRerank-Bored} & 10 & 10 & 10 & 10 & 5 & 0.30 & 1.0 & 0.85 & 10 & 5 & 5 & 1.0 & full \\ \bottomrule
\end{tabular}
}
\end{table}

The hyperparameters used for each of the environments introduced in Section~\ref{sec:env} are detailed in Table~\ref{tab:hyperparam-val}. The hyperparameter values were chosen to reflect the environment-specific characteristics that we highlighted in Table~\ref{tab:env}.

We now describe the hyperparameters used for the different methods.\footnote{With our source code we provide the detailed hyperparameters used for each agent on each environment to facilitate reproducibility.} 
For all RL recommendation agents, we set the discount factor $\gamma$ to 0.0 for static environments (without boredom and influence) and 0.8 for dynamic ones (with boredom and/or influence). Agents are trained for 500,000 steps, where each step corresponds to the agent producing a recommendation ~-- a slate or a single item depending on the environment. The policy learning rate and critic learning rate (for approaches with a critic) were fixed to 0.0003 and 0.01, respectively. Actors and Q-networks are MLPs with a hidden size of 256 at all layers. For REINFORCE agents, the buffer size was set to 100. For SAC-based approaches and HAC, we used a buffer size of $10^6$, a batch size of 32, and a target smoothing coefficient $\tau$ equal to respectively 0.05 and 0.5. In SAC-based agents, we adopted auto-tuning for the entropy regularization coefficient $\alpha$ and we used a single Q-network. We also independently tuned the hyperparameters specific to the HAC approach: we set the learning rate of the behavior loss to 0.00003, the standard deviation for the reparameterization trick to 0.1, the weight for the hyper-actor loss to 0.1, and the dimension of the latent space to 32.

For environments with partial observability (\texttt{SingleItem-PartialObs}, \texttt{SlateTopK-PartialObs}, \texttt{SlateTopK-Uncertain}), we used two types of state encoders commonly used in RL-based recommender systems \citep{Huang2022}: GRU and transformer. The input to the state encoder is a sequence containing for each step the concatenation of click embeddings and item embeddings averaged over the slate. The click and item embeddings are learned independently in the state encoder. The click embedding dimension was set to 2, and the item embedding dimension was set to 16 for the GRU state encoder and 32 for the transformer state encoder. The dimension of the state output by the state encoder was fixed to 32. We used 2 layers (both for the GRU and the transformer), as well as 4 attention heads, a dropout rate of 0.1, and a feedforward dimension of 64 (only for the transformer).

\subsection{Evaluation protocol and metrics}
\label{sec:metric}

For the evaluation, we performed 5 seeded runs for each method on each environment. For each run, we recorded the validation performance on 25 validation episodes every 50,000 training steps. An episode corresponds to a session of $L$ steps where each step corresponds to the agent issuing a recommendation. For every episode, we sample a new random user embedding following the procedure described in Section~\ref{sec:embeddings}. For that reason, validation users are distinct from training users, ensuring no leakage between training and validation. We also used different seeds during hyperparameter tuning (detailed in Section \ref{sec:hyperparam}) and evaluation, in order to have different validation users in these two phases and thus avoid the situation where methods would be specifically optimized on the set of users sampled for tuning.

We considered two metrics in our evaluation. The first one is the return (i.e., the cumulated reward over an episode), averaged over the 25 validation episodes. This metric ranges from 0 to $L \times S$ (i.e., 100 for \texttt{SingleItem} environments and 1000 for \texttt{SlateTopK} environments), which corresponds to the case where the user clicked on all the items presented to them. A higher return indicates more clicks from the user across the episode and thus higher quality recommendation from the agent. On the environments that include a boredom mechanism (\texttt{SingleItem-BoredInf}, \texttt{SlateTopK-Bored}, \texttt{SlateTopK-BoredInf}, \texttt{SlateTopK-PartialObs}, \texttt{SlateTopK-Uncertain}), we also report the boredom metric. We define this metric as the number of steps in the episode where the user is bored on at least one topic~-- lower is better. In our churn-an-return boredom setting (see Section~\ref{sec:boredom} for more details), this corresponds to the number of steps where the user embedding is zeroed out and the user cannot click. This metric is important as in order to be successful an agent should be able to balance accurate recommendations (to reach high immediate rewards) and diverse recommendations over time (to avoid triggering boredom and temporary churn).

\section{Results}
\label{experiments}

In this section, we describe the results of the experiments done on single item recommendation (Section~\ref{sec:exp-single}), slate top-K recommendation (Section~\ref{sec:exp-topk}), and slate reranking (Section~\ref{sec:exp-reranking}). Again, we remind the reader that the goal of these experiments is to demonstrate the possibilities and challenges of the environments derived from SARDINE, rather than to create a benchmark for state-of-the-art approaches on a limited set of environments. These experiments should be seen as a starting point for researchers and practitioners to further investigate the specific scenarios and approaches of their interest.

\subsection{Experiments on single item recommendation}
\label{sec:exp-single}

\begin{figure*}[t]
\centering
    \begin{subfigure}{.495\textwidth}
    \includegraphics[width = \textwidth]{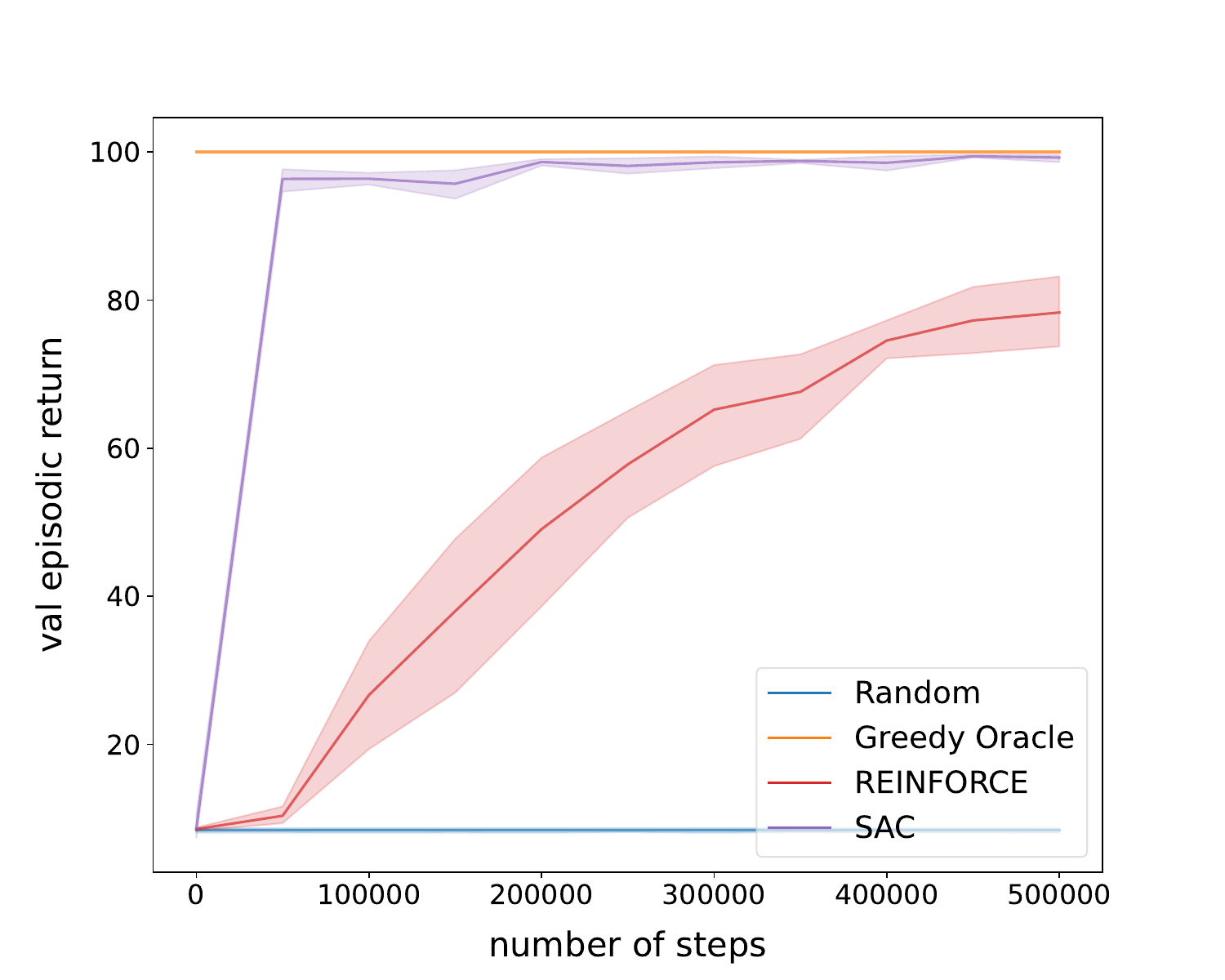} 
    \caption{Return ($\uparrow$) on \texttt{SingleItem-Static}}
    \label{fig:SingleItem-Static-return}
    \end{subfigure}
    \hfill
    \begin{subfigure}{.495\textwidth}
    \includegraphics[width = \textwidth]{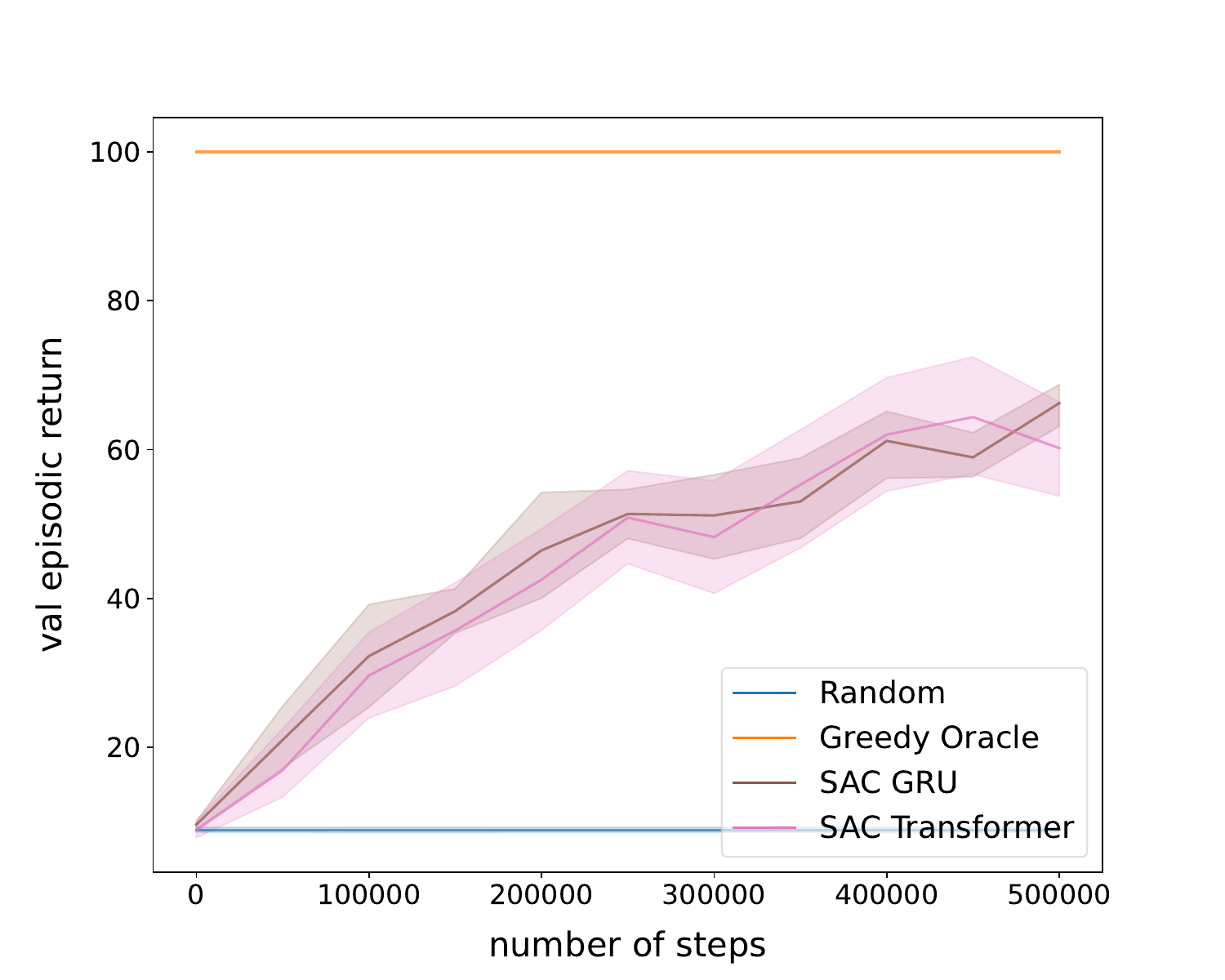} 
    \caption{Return ($\uparrow$) on \texttt{SingleItem-PartialObs}}
    \label{fig:SingleItem-PartialObs-return}
    \end{subfigure}
    \vfill
    \begin{subfigure}{.495\textwidth}
    \includegraphics[width = \textwidth]{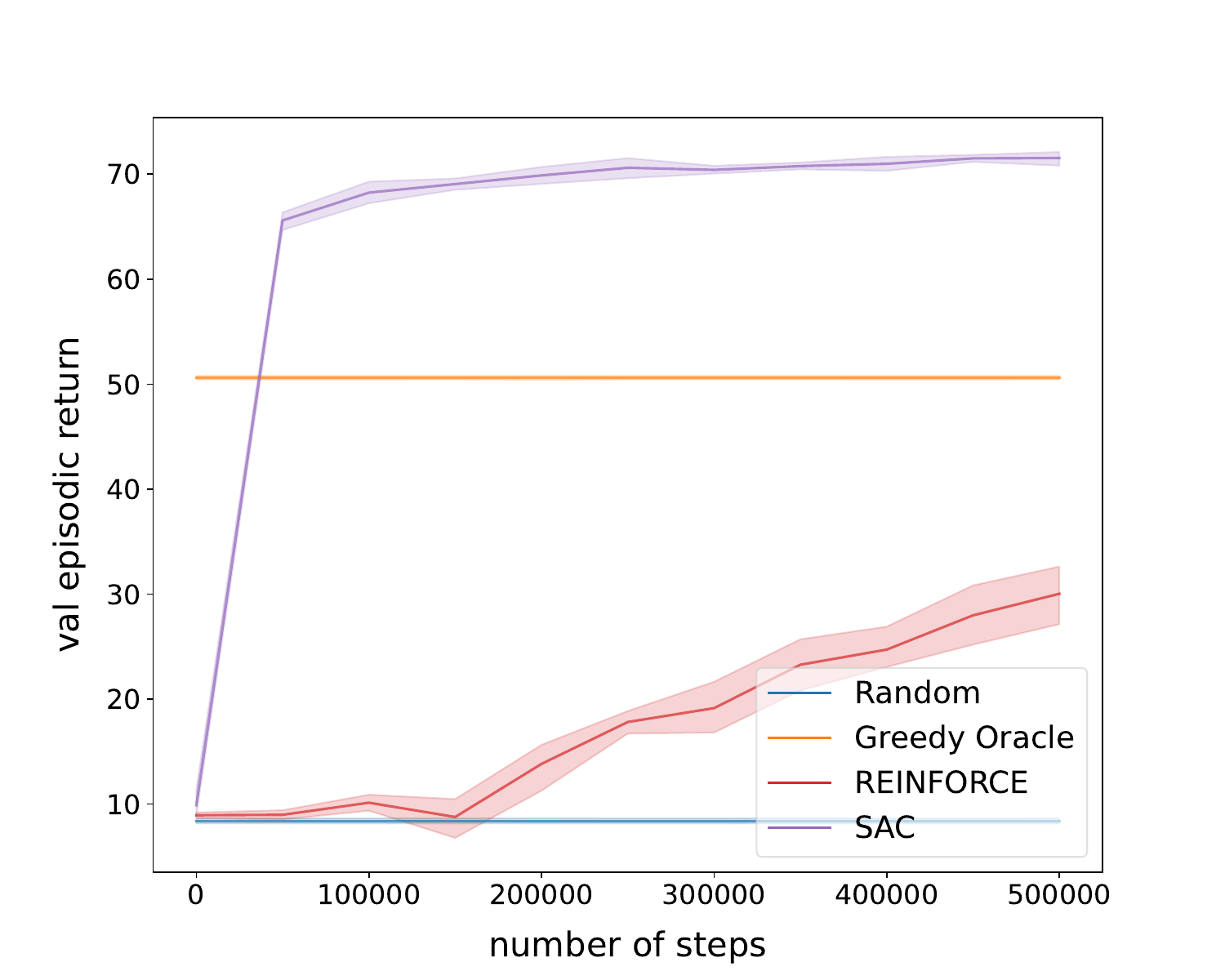} 
    \caption{Return ($\uparrow$) on \texttt{SingleItem-BoredInf}}
    \label{fig:SingleItem-BoredInf-return}
    \end{subfigure}
    \hfill
    \begin{subfigure}{.495\textwidth}
    \includegraphics[width = \textwidth]{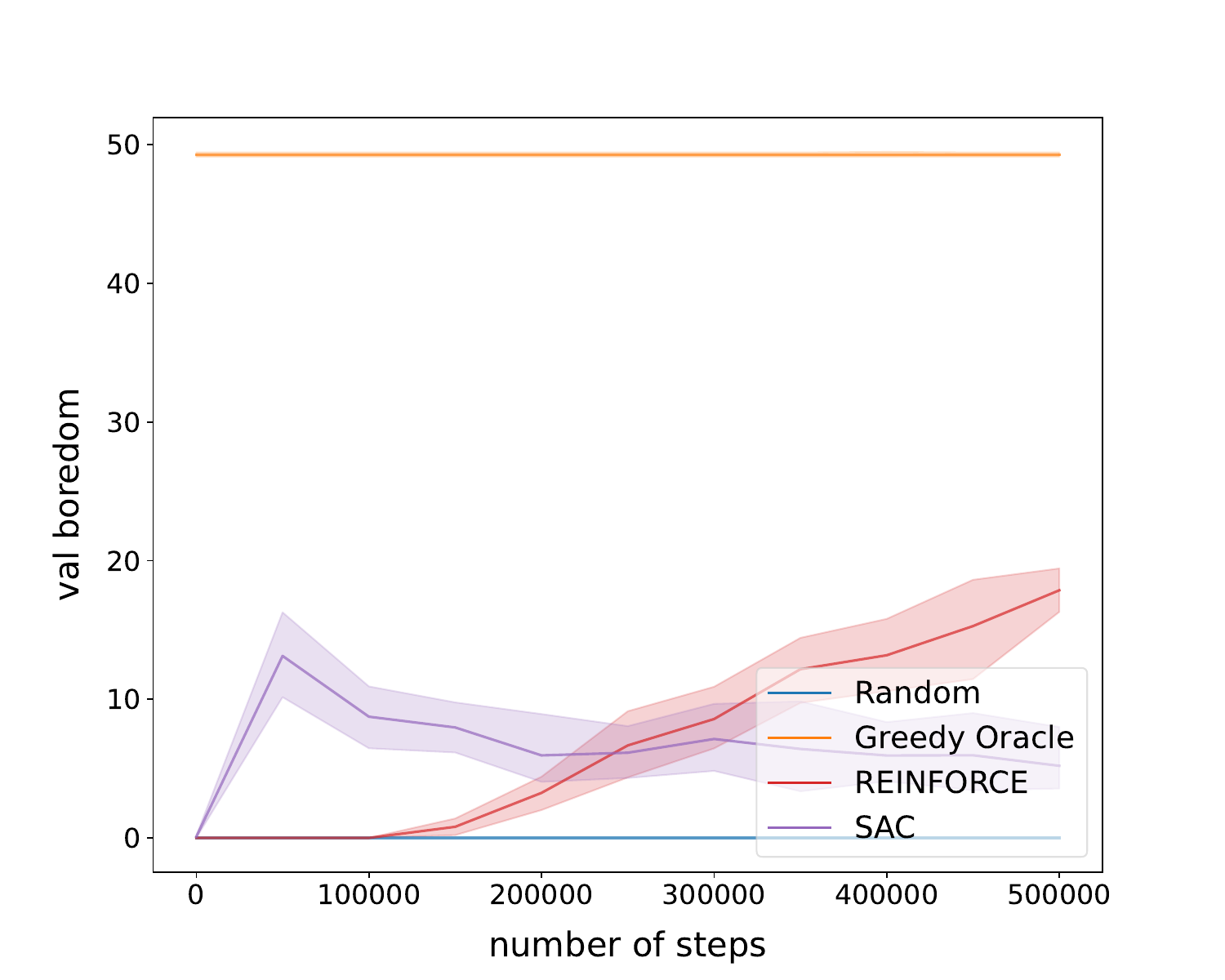} 
    \caption{Boredom ($\downarrow$) on \texttt{SingleItem-BoredInf}}
    \label{fig:SingleItem-BoredInf-boredom}
    \end{subfigure}
\vspace{0.1cm}
\caption{Results on the \texttt{SingleItem-Static} (\ref{fig:SingleItem-Static-return}), \texttt{SingleItem-PartialObs} (\ref{fig:SingleItem-PartialObs-return}), and \texttt{SingleItem-BoredInf} (\ref{fig:SingleItem-BoredInf-return}, \ref{fig:SingleItem-BoredInf-boredom}) environments. The colored envelope surrounding lines indicates the 95\% confidence interval around the mean computed from 5 seeded runs. Boredom results are not shown for \texttt{SingleItem-Static} and \texttt{SingleItem-PartialObs} as these static environments do not include a boredom component and thus all methods have a default boredom of 0.}
\label{fig:SingleItem}       
\end{figure*}

We performed experiments on three \texttt{SingleItem} environments whose characteristics are recalled below (see Section~\ref{sec:env} for a more detailed description). In \texttt{SingleItem-Static}, we consider an easy, static recommendation scenario in a fully observable setting and with low uncertainty in order to validate that learned agents can reach optimal performance. \texttt{SingleItem-BoredInf} adds boredom and influence mechanisms to \texttt{SingleItem-Static}, increasing the difficulty of the environment. For \texttt{SingleItem-PartialObs}, we start as well from \texttt{SingleItem-Static} but change it to a POMDP setting. The results on the \texttt{SingleItem} environments are given in Fig.~\ref{fig:SingleItem} and discussed below.

\paragraph{\bf \texttt{SingleItem-Static}.} The validation return over the different training steps on \texttt{SingleItem-Static} is plotted in Fig.~\ref{fig:SingleItem-Static-return}. In this experiment we compared SAC and REINFORCE agents against the Greedy Oracle and Random baselines. In this specific setting, the Greedy Oracle is by design optimal due to the absence of long-term mechanisms (boredom or influence). It is therefore not surprising that the Greedy Oracle achieves a return of 100, meaning that all recommended items in the session have been clicked. However, it is interesting to note that among the learned agents, SAC fares much better than REINFORCE. Indeed, the former is able to reach optimal performance (or very close to it) after only 200,000 steps, whereas the latter struggles to close the gap. This difference might be explained by the fact that SAC exploits the ideal item embeddings whereas REINFORCE simply selects actions through a softmax over items. Additionally, SAC is generally a better performing RL agent than REINFORCE in most cases, due to a better bias-variance trade-off. Nonetheless, it is reassuring to see that in this simple environment, a learned agent such as SAC is able to easily find the optimal policy.

\paragraph{\bf \texttt{SingleItem-BoredInf}.} We now turn to the more challenging \texttt{SingleItem-BoredInf} environment which includes boredom and influence mechanisms. The results according to the return and boredom metrics are illustrated in Fig.~\ref{fig:SingleItem-BoredInf-return} and \ref{fig:SingleItem-BoredInf-boredom}, respectively. This environment corresponds to typical RL-based recommendation in an MDP setting and we expect RL agents to be able to beat a myopic approach such as the Greedy Oracle. Indeed, the Greedy Oracle is no longer optimal here due to the introduction of long-term mechanisms. This is confirmed by the results in the plots, which show that the Greedy Oracle only yields a return of around 50, and a boredom of around 50 as well. This means that for 50\% of the steps the user is in a bored state, and for the remaining 50\% the recommendation leads to a click. Turning to the RL agents, we first see that REINFORCE struggles to learn an effective policy and remains inferior to the Greedy Oracle in terms of return. However, SAC is able to provide high-quality recommendations, with a return close to 70 even after only 50,000 training steps. SAC is also able to recommend diversified items, as its low (and diminishing) boredom confirms. On the other hand, the boredom of REINFORCE increases steadily as return increases, showing that its policy is still in an accuracy improvement stage and is not favoring result diversification.

\paragraph{\bf \texttt{SingleItem-PartialObs}.} This environment is static like \texttt{SingleItem-Static} but we consider here a POMDP setting, i.e., partial state observability. This corresponds to the typical sequential recommendation scenario based on offline feedback, where recommendations have no causal effect on user behavior \citep{offline-eval-RL4REC}. Due to the partial observability, RL agents require a state encoder to convert the observations returned by the environment into a state that can be exploited by the agent. We consider both a GRU and a transformer state encoder, which we test in combination with a SAC agent as it obtained convincing results on \texttt{SingleItem-Static}. The results are shown in Fig.~\ref{fig:SingleItem-PartialObs-return}. Here, the Greedy Oracle is still optimal as it has access to the true user state (and thus is not affected by the POMDP setting). The partial observability does have an impact on the SAC agents, leaving a gap between the 60+ return obtained by these and the 100 return obtained by the Greedy Oracle. This highlights that more research might be needed on state encoders to be able to accurately estimate the true user state in this setting. Comparing the variants of SAC equipped with a GRU and transformer state encoder, we do not notice statistically significant differences between those two in this case.

\subsection{Experiments on slate top-K recommendation}
\label{sec:exp-topk}

\begin{figure*}[t]
\centering
    \begin{subfigure}{.495\textwidth}
    \includegraphics[width = \textwidth]{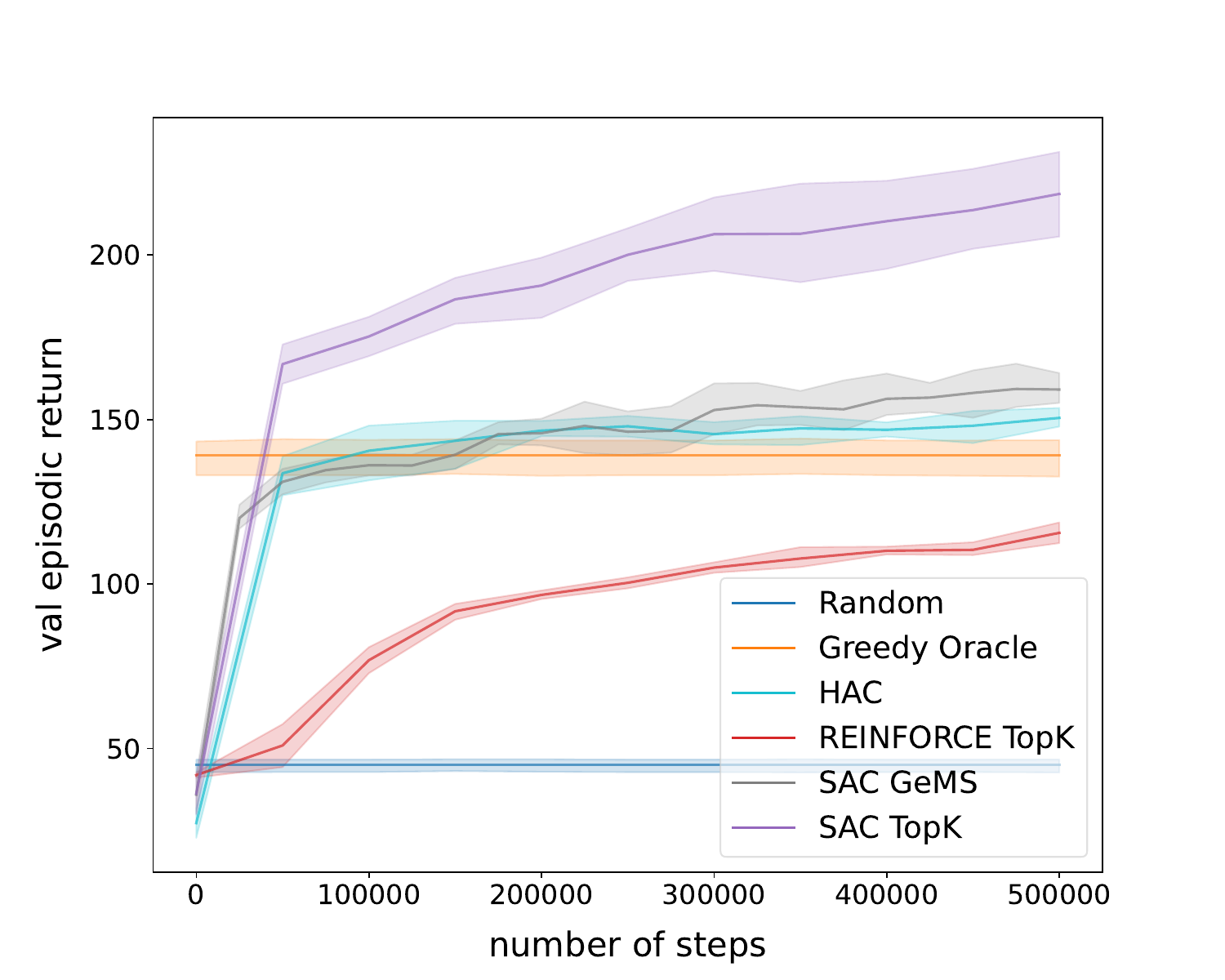} 
    \caption{Return ($\uparrow$) on \texttt{SlateTopK-Bored} (Ideal)}
    \label{fig:SlateTopK-Bored-return}
    \end{subfigure}
    \hfill
    \begin{subfigure}{.495\textwidth}
    \includegraphics[width = \textwidth]{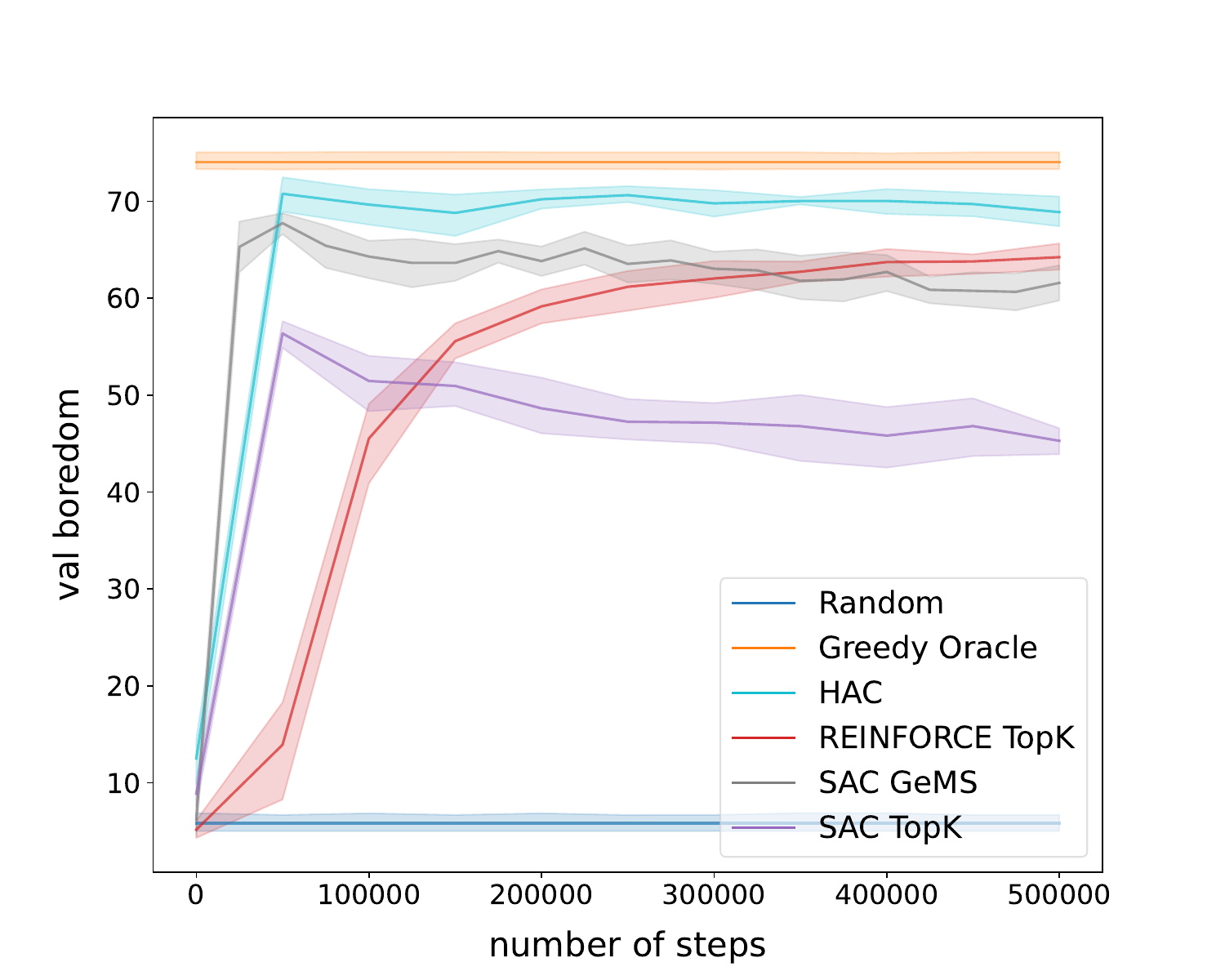} 
    \caption{Boredom ($\downarrow$) on \texttt{SlateTopK-Bored} (Ideal)}
    \label{fig:SlateTopK-Bored-boredom}
    \end{subfigure}
    \vfill
    \begin{subfigure}{.495\textwidth}
    \includegraphics[width = \textwidth]{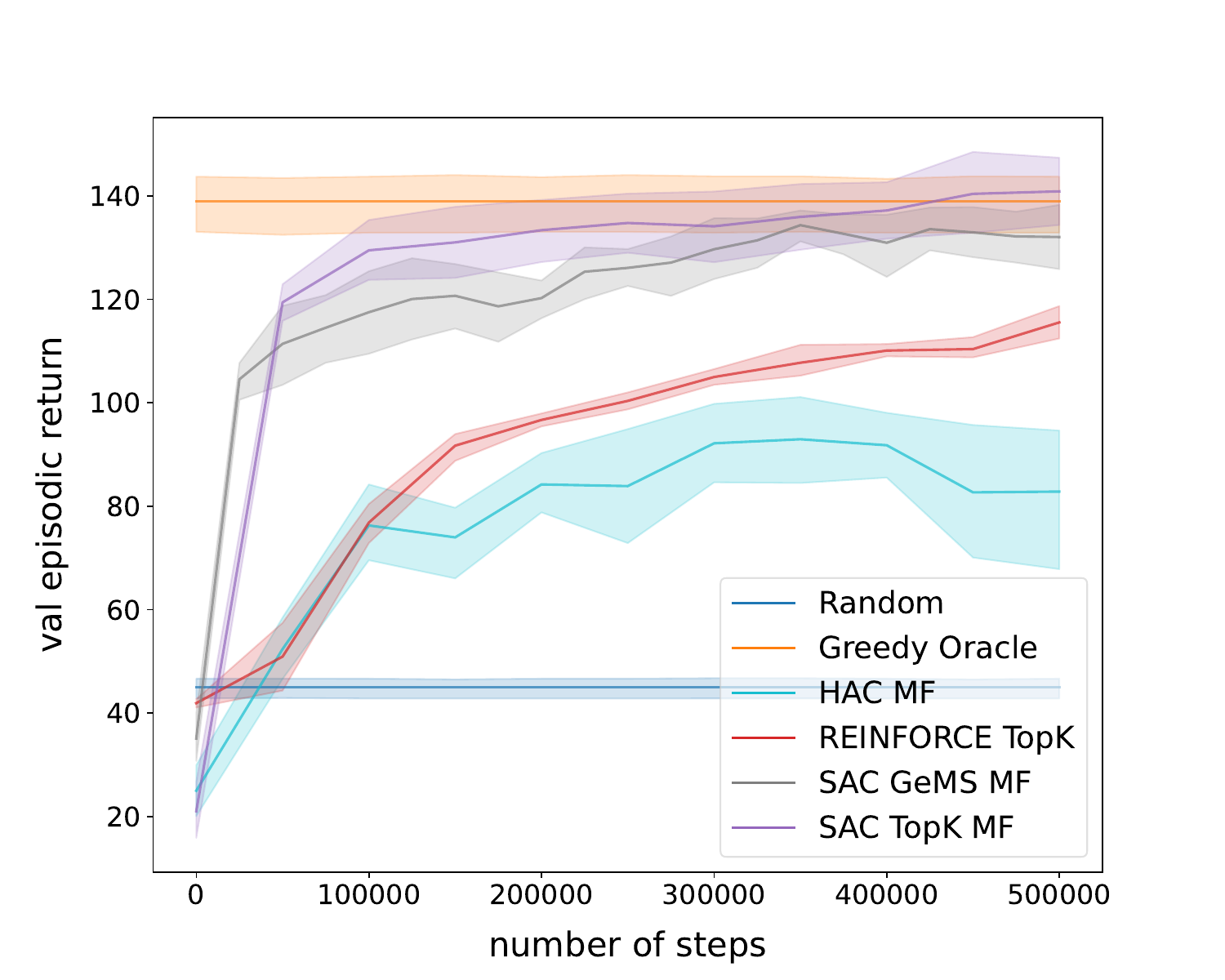} 
    \caption{Return ($\uparrow$) on \texttt{SlateTopK-Bored} (MF)}
    \label{fig:SlateTopK-Bored-MF-return}
    \end{subfigure}
    \hfill
    \begin{subfigure}{.495\textwidth}
    \includegraphics[width = \textwidth]{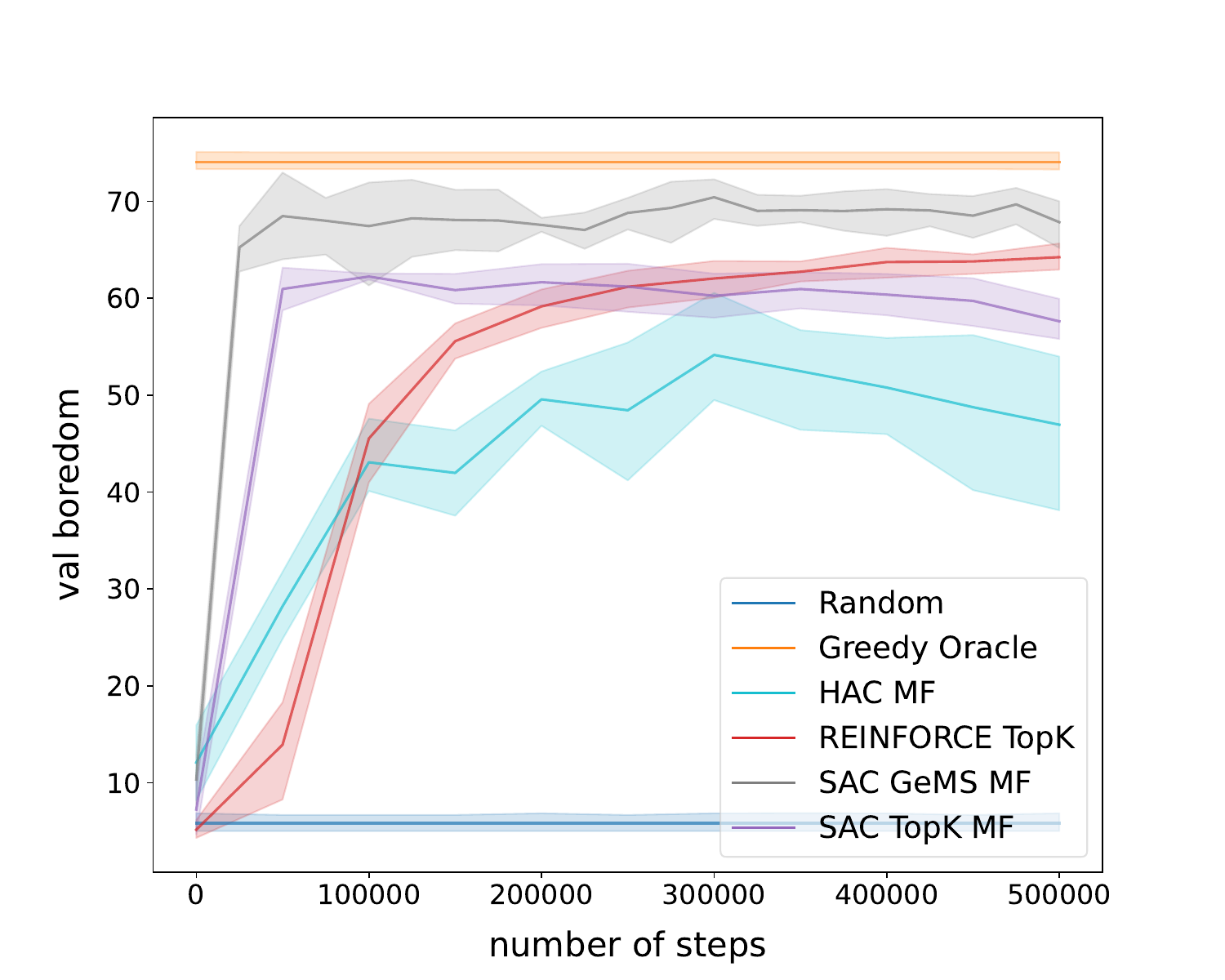} 
    \caption{Boredom ($\downarrow$) on \texttt{SlateTopK-Bored} (MF)}
    \label{fig:SlateTopK-Bored-MF-boredom}
    \end{subfigure}
\vspace{0.1cm}
\caption{Results on the \texttt{SlateTopK-Bored} environment with default, ideal item embeddings (\ref{fig:SlateTopK-Bored-return}, \ref{fig:SlateTopK-Bored-boredom}) and with matrix factorization item embeddings (\ref{fig:SlateTopK-Bored-MF-return}, \ref{fig:SlateTopK-Bored-MF-boredom}). The colored envelope surrounding lines indicates the 95\% confidence interval around the mean computed from 5 seeded runs. Some approaches keep the same performance across the two settings as they either do not rely on item embeddings (Random, REINFORCE Top-K) or are an oracle baseline and only make sense with ideal item embeddings (Greedy Oracle).}
\label{fig:SlateTopK-Bored}       
\end{figure*}

\begin{figure*}[t]
\centering
    \begin{subfigure}{.495\textwidth}
    \includegraphics[width = \textwidth]{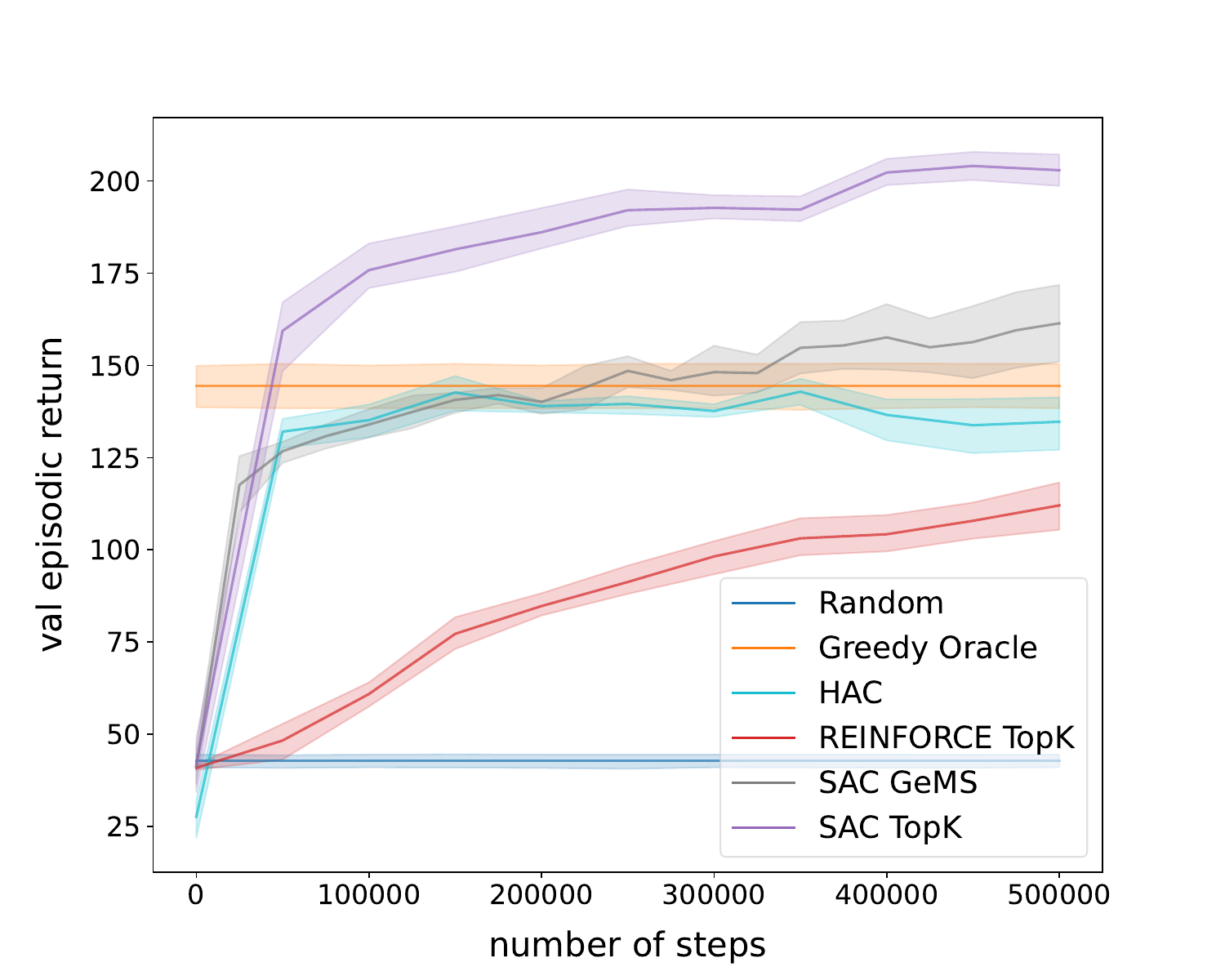} 
    \caption{Return ($\uparrow$)}
    \label{fig:SlateTopK-BoredInf-return}
    \end{subfigure}
    \hfill
    \begin{subfigure}{.495\textwidth}
    \includegraphics[width = \textwidth]{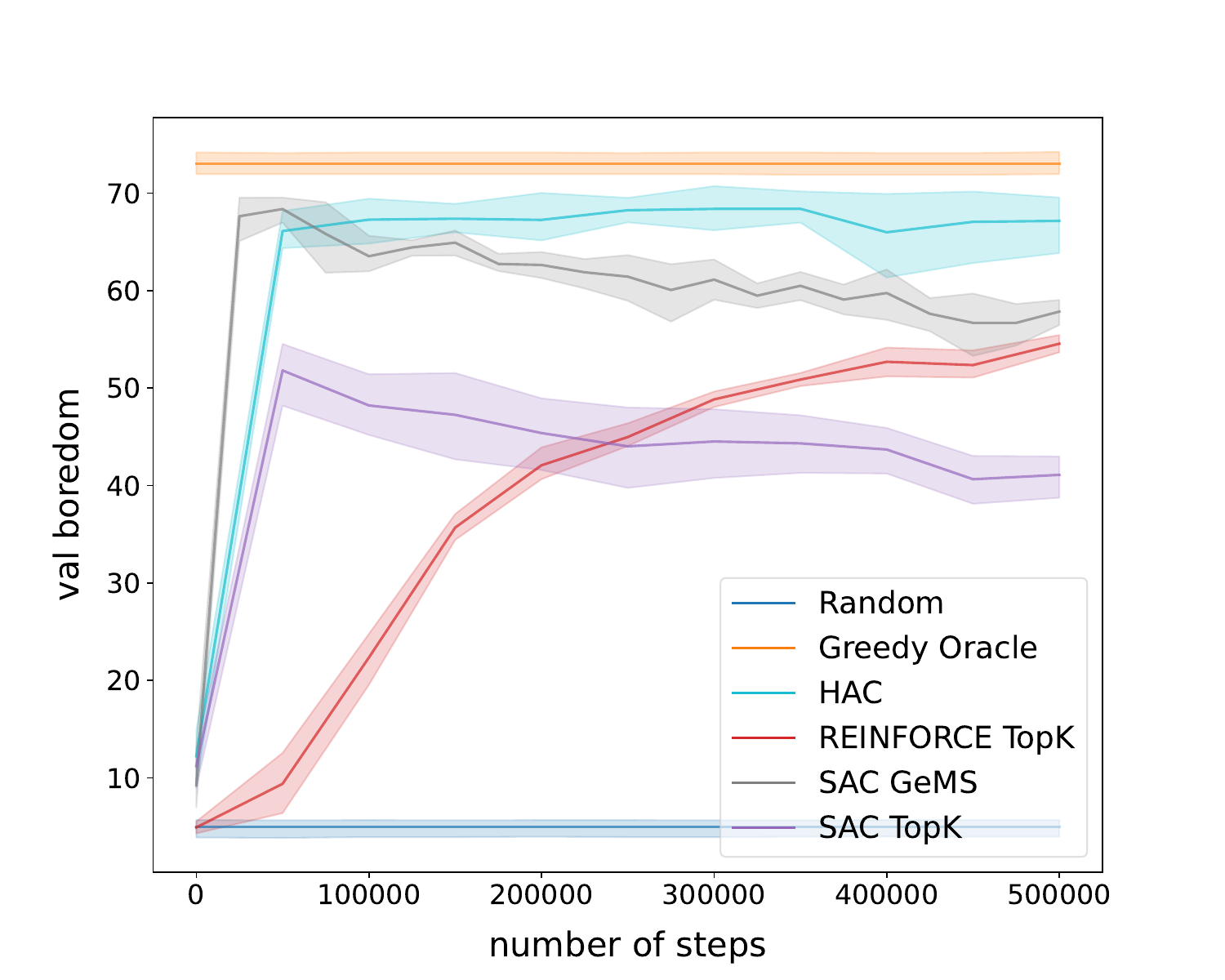} 
    \caption{Boredom ($\downarrow$)}
    \label{fig:SlateTopK-BoredInf-boredom}
    \end{subfigure}
\vspace{0.1cm}
\caption{Results on the \texttt{SlateTopK-BoredInf} environment. The colored envelope surrounding lines indicates the 95\% confidence interval around the mean computed from 5 seeded runs.}
\label{fig:SlateTopK-BoredInf}       
\end{figure*}

\begin{figure*}[t]
\centering
    \begin{subfigure}{.495\textwidth}
    \includegraphics[width = \textwidth]{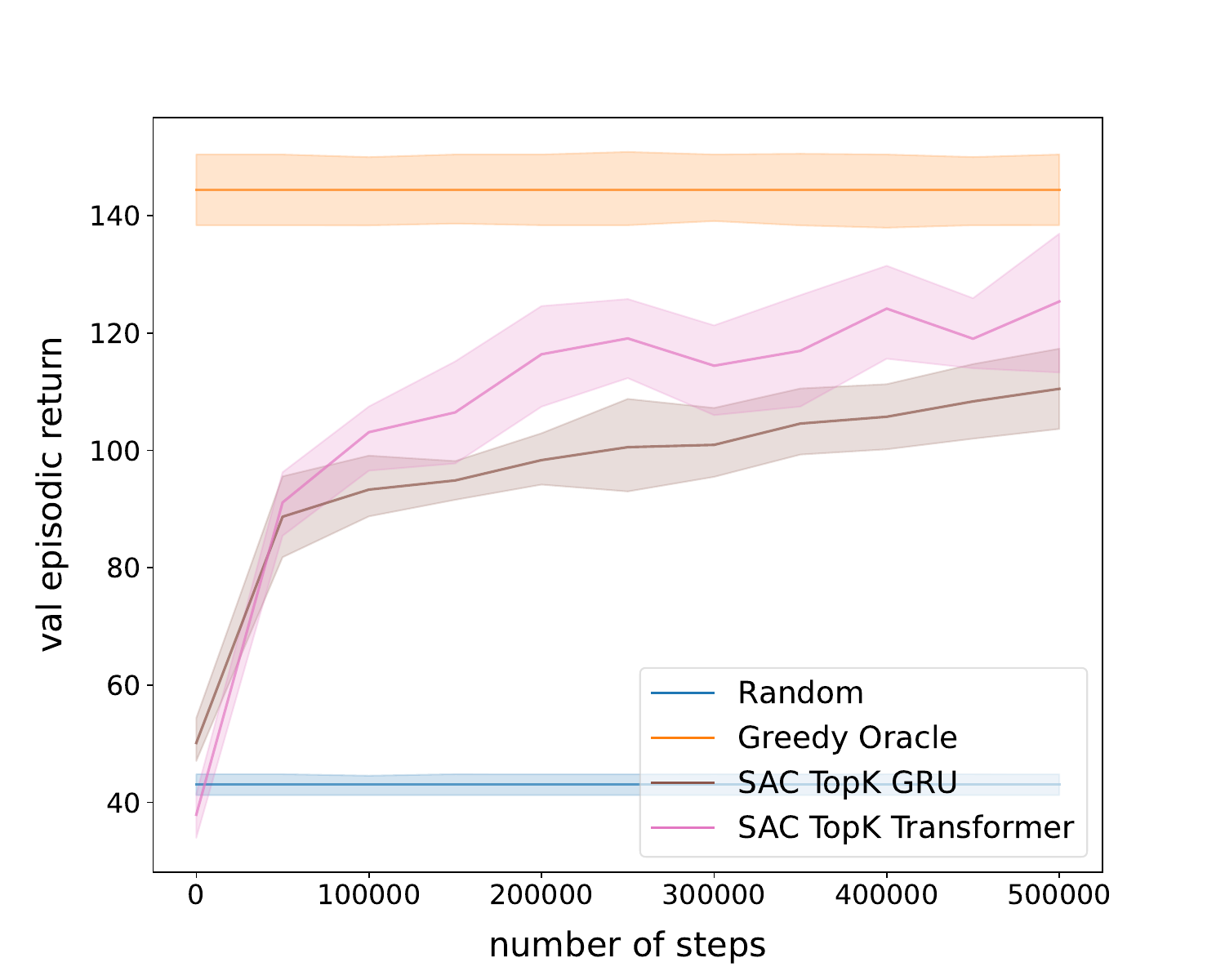} 
    \caption{\texttt{SlateTopK-PartialObs}, low uncertainty}
    \label{fig:SlateTopK-PartialObs-return}
    \end{subfigure}
    \hfill
    \begin{subfigure}{.495\textwidth}
    \includegraphics[width = \textwidth]{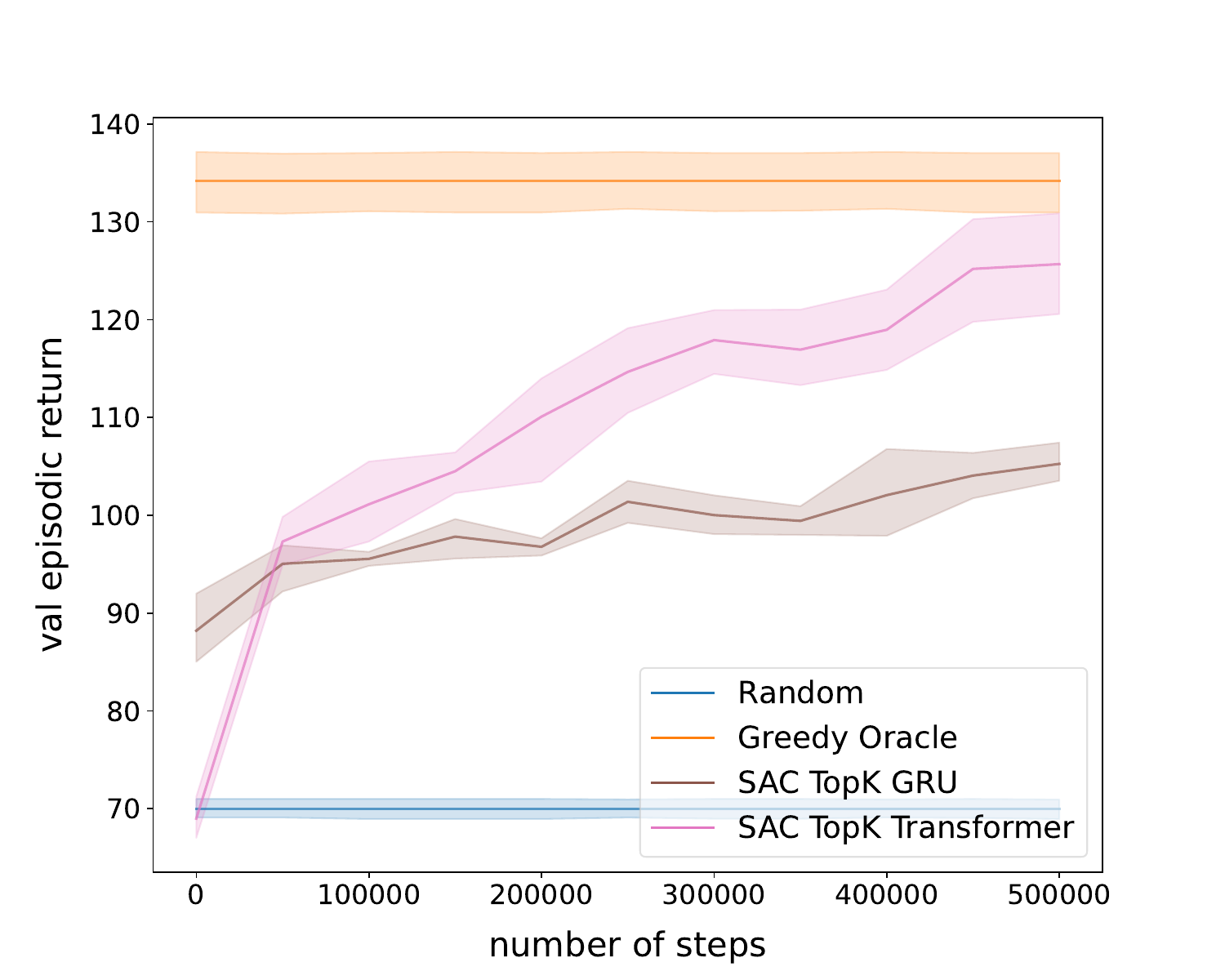} 
    \caption{\texttt{SlateTopK-Uncertain}, medium uncertainty}
    \label{fig:SlateTopK-Uncertain10-return}
    \end{subfigure}
    \vfill
    \begin{subfigure}{.495\textwidth}
    \includegraphics[width = \textwidth]{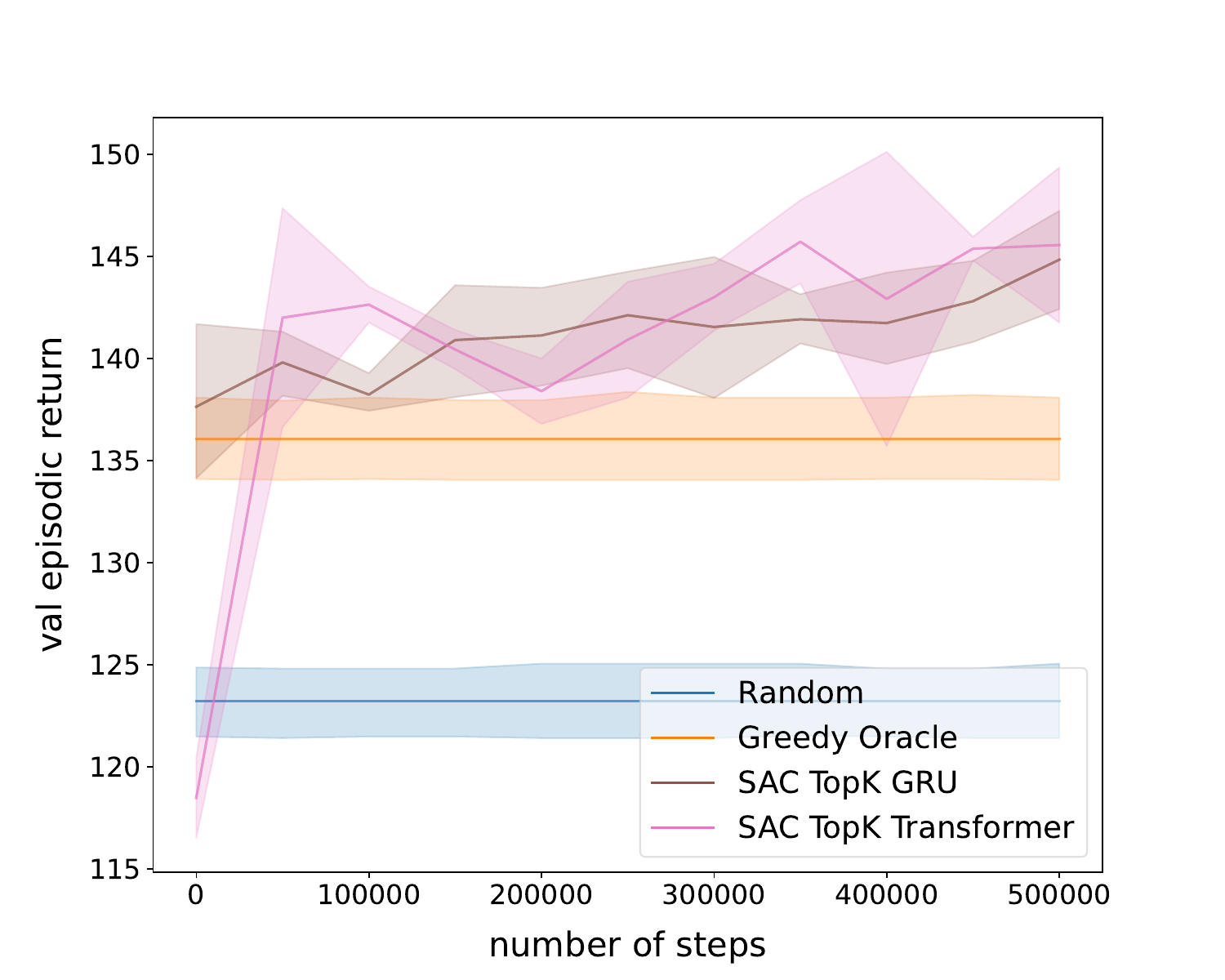} 
    \caption{\texttt{SlateTopK-Uncertain}, high uncertainty}
    \label{fig:SlateTopK-Uncertain5-return}
    \end{subfigure}
    \hfill
    \begin{subfigure}{.495\textwidth}
    \includegraphics[width = \textwidth]{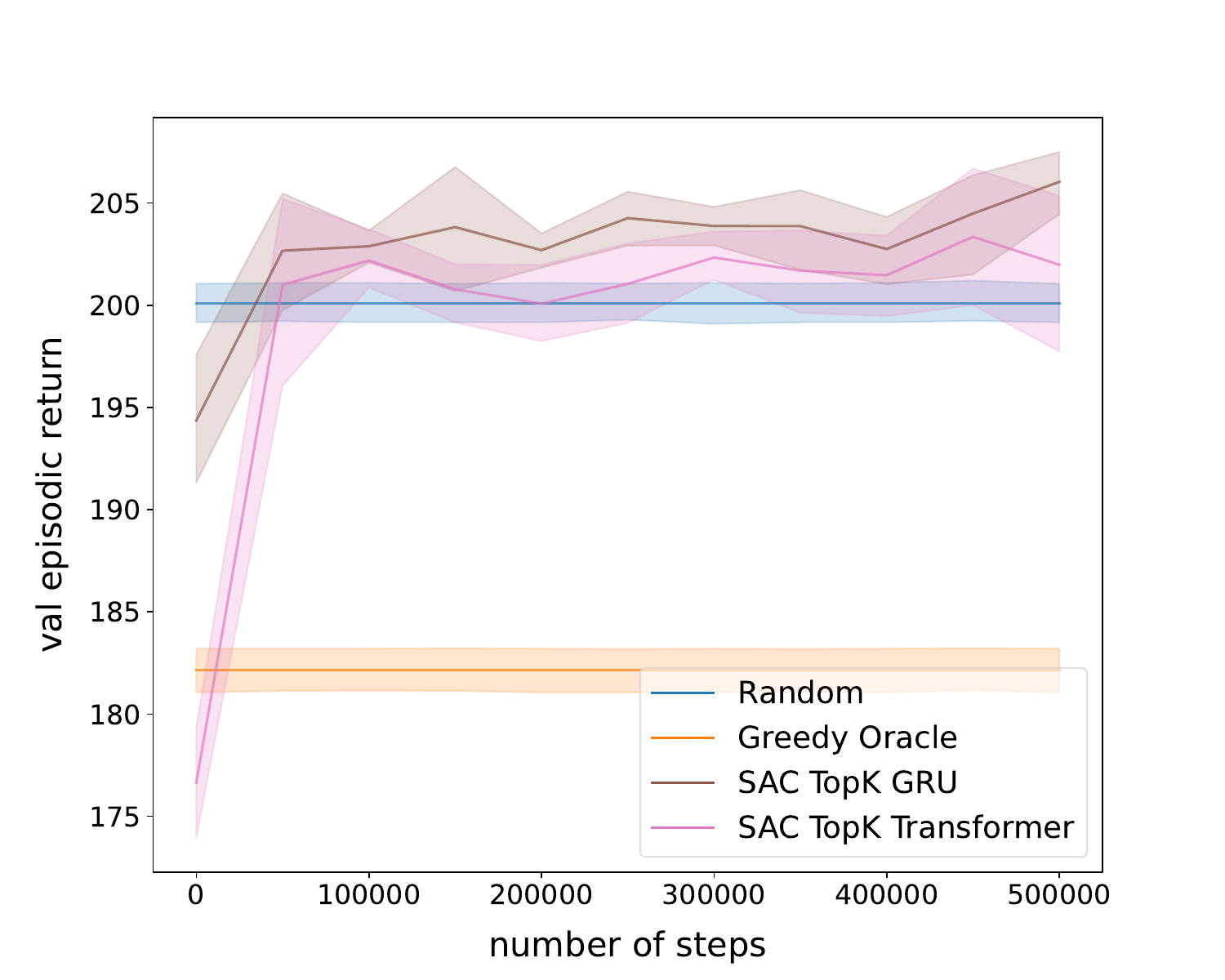} 
    \caption{\texttt{SlateTopK-Uncertain}, very high uncertainty}
    \label{fig:SlateTopK-Uncertain2-return}
    \end{subfigure}
\vspace{0.1cm}
\caption{Results in terms of return ($\uparrow$) on the \texttt{SlateTopK-PartialObs} (\ref{fig:SlateTopK-PartialObs-return}) and \texttt{SlateTopK-Uncertain} (\ref{fig:SlateTopK-Uncertain10-return}, \ref{fig:SlateTopK-Uncertain5-return}, \ref{fig:SlateTopK-Uncertain2-return}) environments. The click uncertainty degree varies from low (\ref{fig:SlateTopK-PartialObs-return}), medium (\ref{fig:SlateTopK-Uncertain10-return}), high (\ref{fig:SlateTopK-Uncertain5-return}) to very high (\ref{fig:SlateTopK-Uncertain2-return}), corresponding to a scale hyperparameter $\lambda$ in the relevance function equal to 100, 10, 5, and 2, respectively (see Section~\ref{sec:click} for more details). The colored envelope surrounding lines indicates the 95\% confidence interval around the mean computed from 5 seeded runs.}
\label{fig:SlateTopK-Uncertain-return}  
\end{figure*}

\begin{figure*}[t]
\centering
    \begin{subfigure}{.495\textwidth}
    \includegraphics[width = \textwidth]{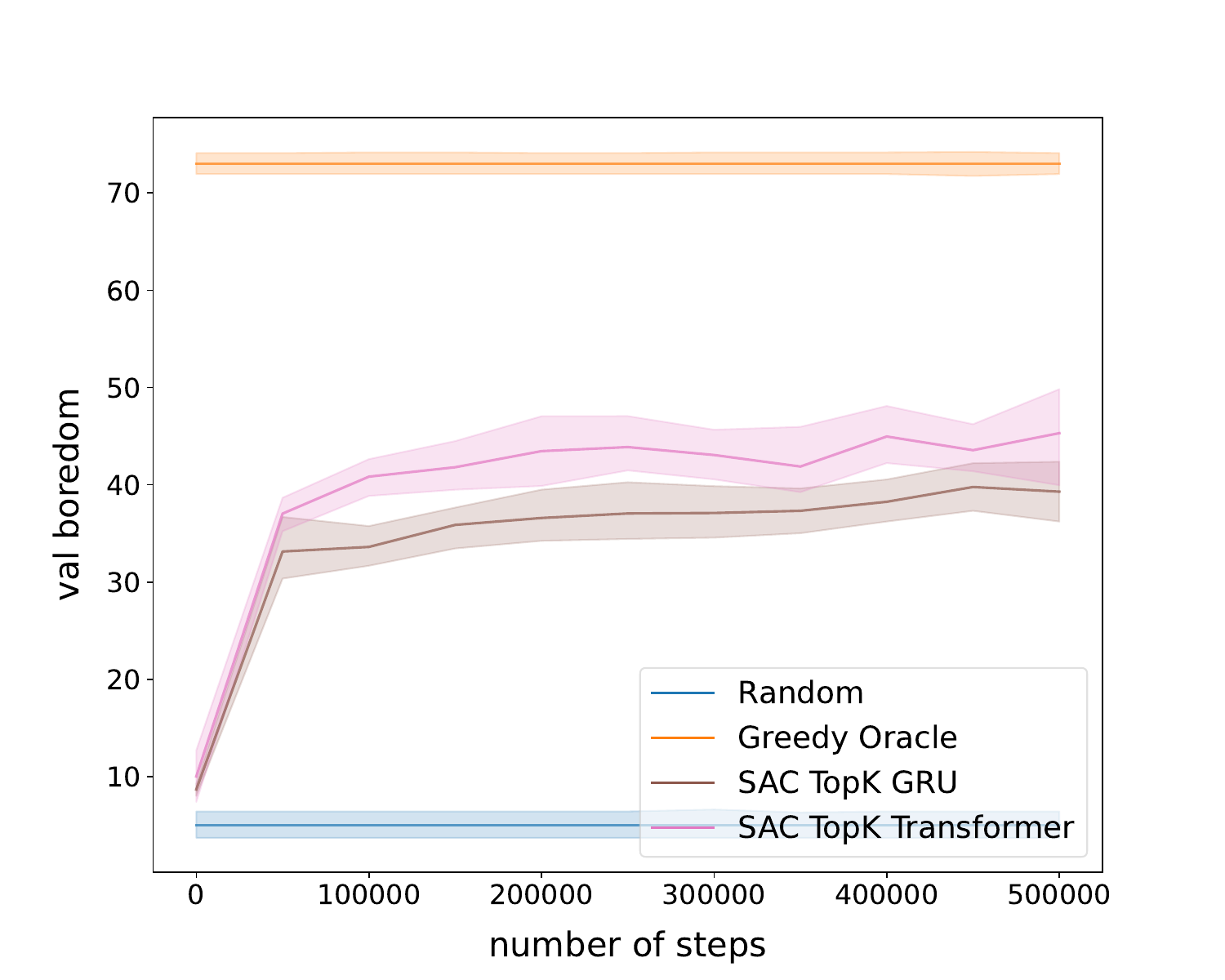} 
    \caption{\texttt{SlateTopK-PartialObs}, low uncertainty}
    \label{fig:SlateTopK-PartialObs-boredom}
    \end{subfigure}
    \hfill
    \begin{subfigure}{.495\textwidth}
    \includegraphics[width = \textwidth]{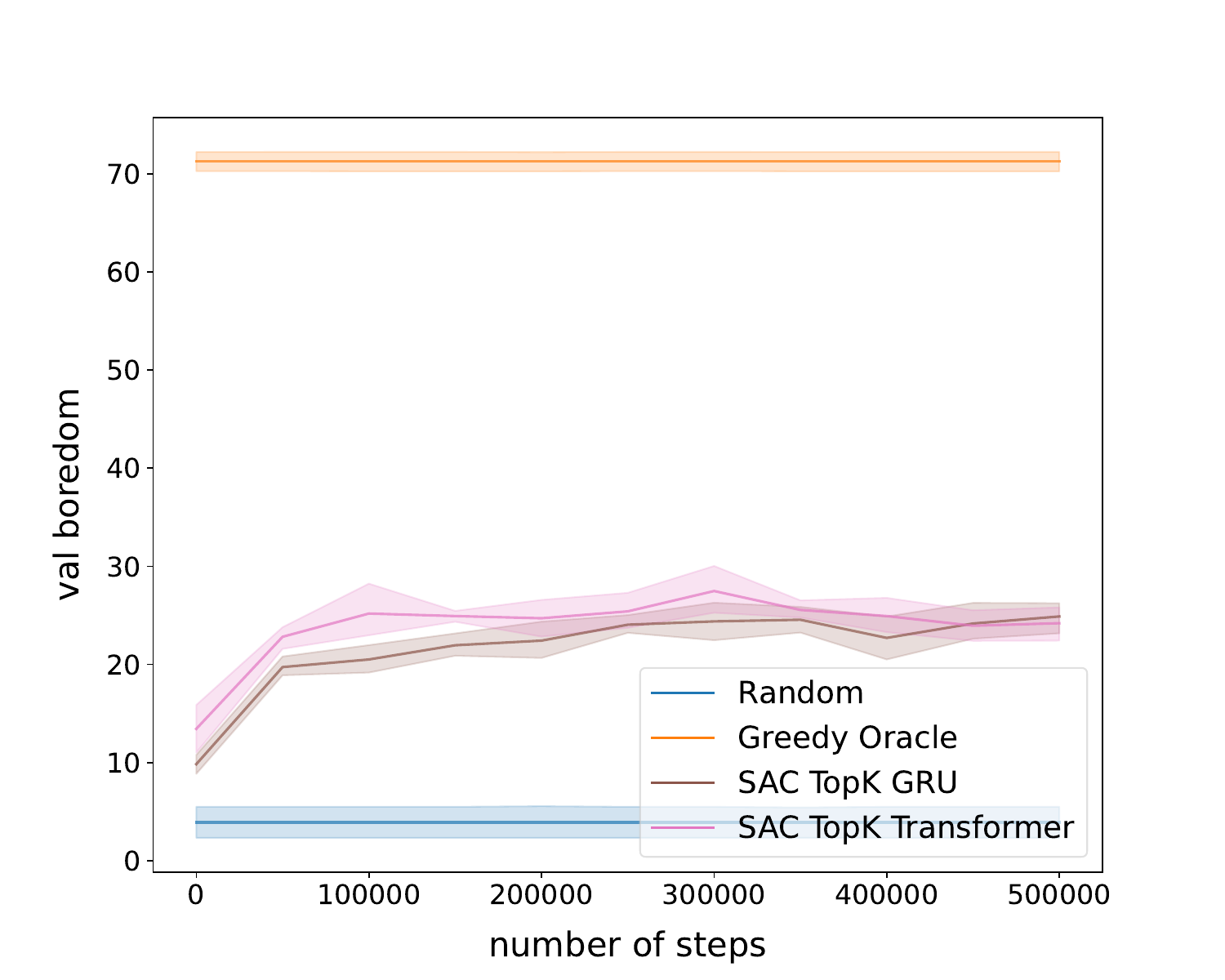} 
    \caption{\texttt{SlateTopK-Uncertain}, medium uncertainty}
    \label{fig:SlateTopK-Uncertain10-boredom}
    \end{subfigure}
    \vfill
    \begin{subfigure}{.495\textwidth}
    \includegraphics[width = \textwidth]{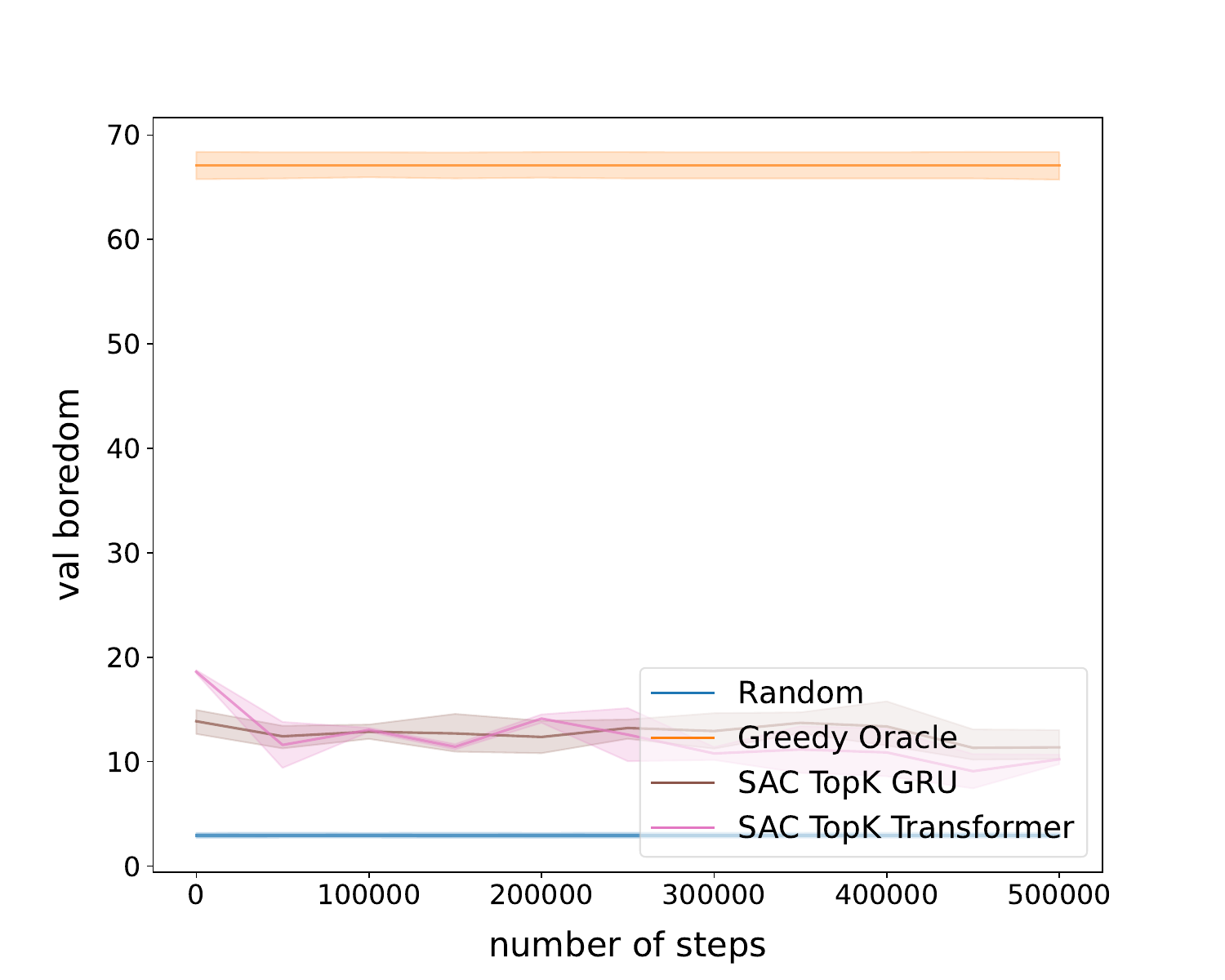} 
    \caption{\texttt{SlateTopK-Uncertain}, high uncertainty}
    \label{fig:SlateTopK-Uncertain5-boredom}
    \end{subfigure}
    \hfill
    \begin{subfigure}{.495\textwidth}
    \includegraphics[width = \textwidth]{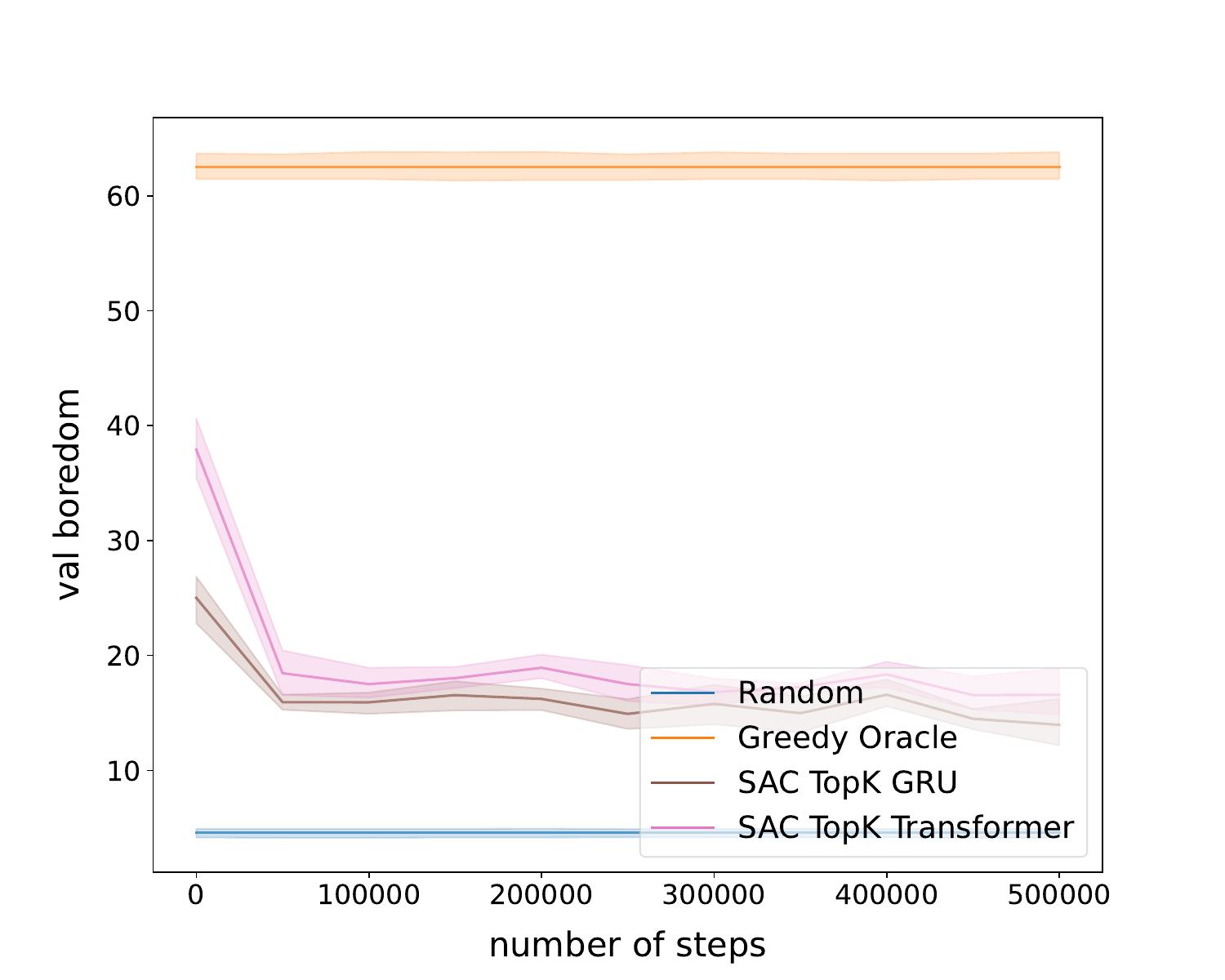} 
    \caption{\texttt{SlateTopK-Uncertain}, very high uncertainty}
    \label{fig:SlateTopK-Uncertain2-boredom}
    \end{subfigure}
\vspace{0.1cm}
\caption{Results in terms of boredom ($\downarrow$) on the \texttt{SlateTopK-PartialObs} (\ref{fig:SlateTopK-PartialObs-boredom}) and \texttt{SlateTopK-Uncertain} (\ref{fig:SlateTopK-Uncertain10-boredom}, \ref{fig:SlateTopK-Uncertain5-boredom}, \ref{fig:SlateTopK-Uncertain2-boredom}) environments. The click uncertainty degree varies from low (\ref{fig:SlateTopK-PartialObs-boredom}), medium (\ref{fig:SlateTopK-Uncertain10-boredom}), high (\ref{fig:SlateTopK-Uncertain5-boredom}) to very high (\ref{fig:SlateTopK-Uncertain2-boredom}), corresponding to a scale hyperparameter $\lambda$ in the relevance function equal to 100, 10, 5, and 2, respectively (see Sections~\ref{sec:click} for more details). The colored envelope surrounding lines indicates the 95\% confidence interval around the mean computed from 5 seeded runs.}
\label{fig:SlateTopK-Uncertain-boredom}  
\end{figure*}

We now move on to the experiments performed on slate top-K recommendation, i.e., where the recommendation presented to the user is a list instead of a single item as in \texttt{SingleItem} environments. We studied four \texttt{SlateTopK} environments which are summarized below (and further detailed in Section~\ref{sec:env}). \texttt{SlateTopK-Bored} and \texttt{SlateTopK-BoredInf} are both fully observable environments which require agents to do multi-step reasoning to perform well. The difference between the two is that the former only includes a boredom mechanism, whereas the latter additionally integrates an influence mechanism~-- causing user embeddings to drift based on clicked items and thus making it more difficult to track user interest. We also investigated a partially observable version of \texttt{SlateTopK-BoredInf} through \texttt{SlateTopK-PartialObs}, further increasing the difficulty of the task. Finally, we experimented with various levels of uncertainty in the clicking process through the \texttt{SlateTopK-Uncertain} environments. The results on the \texttt{SlateTopK} environments are shown in Fig.~\ref{fig:SlateTopK-Bored},~\ref{fig:SlateTopK-BoredInf},~\ref{fig:SlateTopK-Uncertain-return}, and~\ref{fig:SlateTopK-Uncertain-boredom}.

\paragraph{\bf \texttt{SlateTopK-Bored}.} The results on the \texttt{SlateTopK-Bored} environment are plotted in Fig.~\ref{fig:SlateTopK-Bored-return} (for the return) and Fig.~\ref{fig:SlateTopK-Bored-boredom} (for the boredom). We compared the Greedy Oracle baseline, which is sub-optimal here, to four learned agents: REINFORCE + Top-K, SAC + Top-K, SAC + GeMS and HAC. Similar to the results observed on \texttt{SingleItem} environments, we see here that REINFORCE + Top-K fails to reach the performance of the Greedy Oracle, while SAC + Top-K beats by a good margin the Greedy Oracle. HAC and SAC + GeMS are also able to beat the oracle baseline, but by a much smaller margin. In terms of boredom, SAC + Top-K is also the winner with a much lower value. We also report the distribution of item relevance scores for the different methods tested here in Appendix~\ref{app:scores}. We hypothesize that the superiority of SAC + Top-K, in particular over SAC + GeMS and HAC, is due to the use of ideal item embeddings in this experiment. Indeed, SAC + Top-K directly uses the item embedding space as action space and thus rely heavily on the quality of item representations. To investigate this hypothesis, we repeated the same experiment as shown in Fig.~\ref{fig:SlateTopK-Bored-return} and Fig.~\ref{fig:SlateTopK-Bored-boredom}, but we replace the ideal item embeddings used by default in our experiments with item embeddings learned by matrix factorization (MF).\footnote{To obtain these embeddings, we used the dataset generated for SAC + GeMS pretraining (described in Section~\ref{sec:baselines}) to train a matrix factorization model with an embedding dimension of 10.} We report the results in Fig.~\ref{fig:SlateTopK-Bored-MF-return} and Fig.~\ref{fig:SlateTopK-Bored-MF-boredom} for the return and boredom metrics, respectively. We observe that changing from ideal to MF embeddings does have a drastic effect on the recommendation performance (measured by the return metric) of SAC + Top-K, which degraded to the level of the Greedy Oracle. The performance of HAC is also greatly impacted~-- its return dropping even below that of REINFORCE Top-K. SAC + GeMS is the approach that underwent the smallest performance drop in comparison to the ideal embedding case. This suggests that this latter approach might overall be more robust to sub-optimality in item embeddings.

In addition to the results described above on \texttt{SlateTopK-Bored}, we also investigated a challenging variant of this environment that is closer to a real-life scenario, where certain topics tend to co-occur and the distribution of topics for items and users is skewed. The setting and the experiments done in this environment are further detailed in Appendix~\ref{app:webtoon}.

\paragraph{\bf \texttt{SlateTopK-BoredInf}.} On the \texttt{SlateTopK-BoredInf} environment, which adds an influence mechanism to \texttt{SlateTopK-Bored} with ideal embeddings, we observe similar trends as this latter environment. The return and boredom results are given in Fig.~\ref{fig:SlateTopK-BoredInf-return} and Fig.~\ref{fig:SlateTopK-BoredInf-boredom}, respectively. One notable difference with \texttt{SlateTopK-Bored}, however, is that in \texttt{SlateTopK-BoredInf} HAC fails to beat the Greedy Oracle and gets a return that is significantly worse than that of SAC + GeMS. This might be explained by the fact that HAC integrates a supervised click prediction loss which may hinder the model performance due to the greater dynamics in the user embedding caused by the influence drift. 

\paragraph{\bf \texttt{SlateTopK-PartialObs}.} The results on the \texttt{SlateTopK-PartialObs} environment, which increases \texttt{SlateTopK-BoredInf}'s challenge with partial observability, are shown in Fig.~\ref{fig:SlateTopK-PartialObs-return} (for return) and Fig.~\ref{fig:SlateTopK-PartialObs-boredom} (for boredom). Given the superior performance of SAC + Top-K on \texttt{SlateTopK-BoredInf}, we focus here on variants of this method based on a GRU or a transformer state encoder. In this setting, we observe that the performance of the transformer variant leads on most training steps to a significant improvement in terms of return over the GRU variant. This result goes in line with previous findings on state encoders for RL-based recommendation~\citep{Huang2022}. However, both SAC + Top-K variants fail to beat the Greedy Oracle baseline, highlighting the difficulty of this environment and showing that additional efforts on the agent and/or state encoder might be needed to achieve high-quality recommendation.

\paragraph{\bf \texttt{SlateTopK-Uncertain}.} Starting from \texttt{SlateTopK-PartialObs}, we varied the level of uncertainty in the clicks through the $\lambda$ scale hyperparameter in the simulator's relevance function. In particular, we compared the setting of \texttt{SlateTopK-PartialObs} with low uncertainty ($\lambda = 100$) to different \texttt{SlateTopK-Uncertain} environments with medium uncertainty ($\lambda = 10$), high uncertainty ($\lambda = 5$) and very high uncertainty ($\lambda = 2$)~-- which we will refer to as \texttt{SlateTopK-Uncertain10}, \texttt{SlateTopK-Uncertain5}, and \texttt{SlateTopK-Uncertain2} for simplicity. The return and boredom results in these environments are illustrated in Fig.~\ref{fig:SlateTopK-Uncertain-return} and Fig.~\ref{fig:SlateTopK-Uncertain-boredom}, respectively. Comparing the return on \texttt{SlateTopK-PartialObs} (Fig.~\ref{fig:SlateTopK-PartialObs-return}) to \texttt{SlateTopK-Uncertain10} (Fig.~\ref{fig:SlateTopK-Uncertain10-return}), we observe that the gap between the SAC + Top-K transformer and GRU variants increases. Indeed, while the overall performance of SAC + Top-K GRU slightly decreases with the uncertainty increase, we see that SAC + Top-K transformer is able to maintain its performance at around 125 at 500,000 steps. This suggests that the transformer state encoder is more robust to a medium level of uncertainty. When we increase the uncertainty to a high level in \texttt{SlateTopK-Uncertain5} (Fig.~\ref{fig:SlateTopK-Uncertain5-return}), we notice that the SAC Top-K variants beat the Greedy Oracle baseline, and that the gap between the Random baseline and the Greedy oracle shrinks. This is explained by the fact that with more stochasticity in the clicking process, less relevant items get more clicks~-- which reduces the advantage of the greedily optimal recommendations from the Greedy Oracle. Clicks on more varied items also means that user boredom is less likely to be triggered, which is confirmed by the comparison of the boredom scores of the SAC Top-K variants across Fig.~\ref{fig:SlateTopK-PartialObs-boredom}, Fig.~\ref{fig:SlateTopK-Uncertain10-boredom}, and Fig.~\ref{fig:SlateTopK-Uncertain5-boredom}. When the uncertainty level is further increased in \texttt{SlateTopK-Uncertain2}, we observe that the environment rewards random recommendations more than the accurate recommendations from the Greedy Oracle, as shown in Fig.~\ref{fig:SlateTopK-Uncertain2-return}. This is, again, explained by the fact that less relevant items lead to a click probability similar to that of relevant items. In this setting, the SAC Top-K variants both perform similarly to the Random baseline and are thus learning to favor more diverse recommendations over accurate ones. 

\subsection{Experiments on slate reranking}
\label{sec:exp-reranking}

With this last set of experiments, we explore the reranking task using two \texttt{SlateRerank} environments: the static \texttt{SlateRerank-Static} and the interactive, multi-step \texttt{SlateRerank-Bored}. We conduct experiments on click modeling according to the following protocol: \begin{enumerate*}[label=(\roman*)]
\item we generate a dataset of interactions using a certain logging policy, \item we train a click model on the generated dataset, and \item we rerank the items by decreasing amount of relevance, according to the model. \end{enumerate*} For both environments, we use the reverse-oracle policy, i.e., the policy that orders items by \emph{increasing} order of relevance, as the logging policy. It therefore generates a dataset that contains substantial spurious correlations due to position bias. We report the observed return when applying the reranking methods in the live environment in Table~\ref{tab:slate-rerank}.

\begin{table}[t]
    \centering
    \caption{Online return obtained by debiasing the logged data on the reranking environments. Averaged over five seeded runs. }
    \label{tab:slate-rerank}
    \begin{tabular}{l cc}
        \toprule 
        Method & \texttt{SlateRerank-Static} & \texttt{SlateRerank-Bored}\\
        \midrule
        Greedy Oracle & 21.45 & 13.69 \\
        Reverse Oracle & \phantom{0}8.82 & \phantom{0}8.47\\
        dCTR & \phantom{0}9.28 & \phantom{0}8.97\\
        PBM & 21.17 & 13.14 \\
        Online SAC + Top-K & 19.01 & 14.82\\
        \bottomrule
    \end{tabular}
\end{table}

\paragraph{\bf \texttt{SlateRerank-Static}.} On the static environment, the Greedy Oracle policy is the optimal policy, while the Reverse Oracle yields minimal return. We can indeed first verify in Table~\ref{tab:slate-rerank} that an online-trained SAC + Top-K, as in Section~\ref{sec:exp-topk}, even with full observability of user state and ideal item embeddings, does not beat the greedy oracle. 

Secondly, we can see that a \acf{PBM} correctly identifies the biases in the logged data and almost reaches the performance of the oracle policy, while the naive \acf{dCTR} model fails to do so, and barely improves on the Reverse Oracle policy it was trained on. This result does not come as a surprise since the underlying user click model in the \texttt{SlateRerank} environments is also a position-based model. We can nonetheless verify that the learned propensities, i.e., observation probabilities at each rank, match the true propensities of the simulator. We therefore compute the mean-squared error (MSE) of the normalized propensities, i.e., where the probability of observation at the first position is set to 1, and we find that the learned PBM's propensities have an MSE of $0.570$. That indicates that despite documents being correctly ordered, the learned model does not fully match the underlying model.

The experiments on \texttt{SlateRerank-Static} call for further experiments with different underlying user click models and candidate click models, e.g. as done in~\citep{cmip}, so as to investigate the performance of click models under the more realistic settings of model mismatch. While we leave this for readers to experiment with, we turn to another natural extension that is, to the best of our knowledge, unexplored, and that SARDINE enables.

\paragraph{\bf \texttt{SlateRerank-Bored}.} In this interactive environment, the Greedy Oracle policy is not optimal anymore, because the agent must trade off the accuracy and diversity of the most-exposed topics. Indeed, we can see in Table~\ref{tab:slate-rerank} that an online-trained SAC + Top-K agent beats the Greedy Oracle. This environment therefore constitutes a testbed for (offline) RL agents with biased feedback, and notably the combination of RL and click modeling.

Another important difference that comes with this interactive environment is the fact that the logged data may appear relatively noisier to a click model, as the click/skip feedback can be explained by something else than relevance and position: the boredom. While the boredom information is contained in the ideal user state we use for click model training and the model should therefore in theory be able to correctly identify biases, we expect the training process to be harder. We observe what seems to be a slight degradation of relative performance, compared to the static environment. Indeed, while the PBM managed to fill $98$\% of the gap between the logging policy and the oracle policy on the \texttt{Static} environment, it only fills $89$\% of the gap on the \texttt{Bored} environment. But the extent of the degradation is most apparent when we compare the propensities learned by the model in this new dynamic environment. The MSE now increases to $0.915$, compared to $0.570$ in the static case. This suggests that using the learned propensities of the model in downstream tasks, e.g., counterfactual learning-to-rank, fairness or reinforcement learning, is likely to lead to imperfect and biased policies. 

Effectively using the user behavior learned by click models in a dynamic and interactive environment with, e.g., reinforcement learning, including when the learned variables are imperfect, is to the best of our knowledge still an open question. Our proposed simulator offers the possibility to study this topic in an interpretable and controllable way.

\section{Related Work}
\label{related-work}

\begin{table*}[t]
    \centering
    \caption{Comparison of the proposed SARDINE to existing recommendation simulators. \cmark\ indicates that the research topic is addressed by the simulator and $\mathbf{\sim}$ that it is partially addressed. Our assessment of whether the simulators fulfill the specifications is graded according to \{ $-$, $\pm$, $+$ \}. Overall, we find that only RecSim~\citep{recsim} addresses the research agenda that \acs{SARDINE} targets, but that it does not fulfill our specifications for such a simulator.}
    \vspace{0.1cm}
    \begin{tabular}{l cccc cccc}
    \toprule
        \multicolumn{1}{c}{ } & \multicolumn{4}{c}{\textbf{Research agenda \ref{sec:research-agenda}}} & \multicolumn{3}{c}{\textbf{Specifications \ref{specifications}}} \\
        \cmidrule(r){2-5}
        \cmidrule{6-8}
        Simulator & RT1 & RT2 & RT3 & RT4 & Interpretability & Effect isolation & Configurability \\
         \midrule
        RecoGym &  & $\mathbf{\sim}$ & \cmark &  & $\pm$ & $+$  & $+$ \\
        MARS-Gym &  &  &  &  & $-$ & $-$  & $\pm$\\
        RL4RS & $\mathbf{\sim}$ & $\mathbf{\sim}$ &  & \cmark  & $-$ & $-$  & $-$\\
        RecSim & \cmark & \cmark & $\mathbf{\sim}$ & $\mathbf{\sim}$ & $\pm$ & $-$  & $\pm$  \\
        Virtual-TB & \cmark & & \cmark & & $-$ & $-$ & $-$ \\
        SOFA & & \cmark & & & $+$ & $\pm$ & $+$\\
        OBP & & \cmark & & & $+$ & $+$ & $+$\\
        SARDINE & \cmark & \cmark & \cmark & \cmark & $+$ & $+$  & $+$ \\
    \bottomrule
    \end{tabular}
    \label{tab:related-work}
\end{table*}

In this section we highlight how our work differs from previously published simulators. Considering the research agenda we defined in Section~\ref{sec:research-agenda} as well as our target specifications~\ref{specifications}, we draw a comparison of existing simulators, along with our proposed \acs{SARDINE}, in Table~\ref{tab:related-work}. Note that some of these simulators may have been proposed to target a different research outcome, but we analyze only what we think to be relevant to interactive recommendation research and our corresponding research agenda. Also, we acknowledge that some of the criteria used here are subjective and we try to substantiate our claims as much as possible. 

We now describe related simulators that have been published in recent years, and how they may differ from our objective.

\noindent\textbf{RecoGym~\citep{recogym}} is an e-commerce and advertising simulator where the agent aims to display attractive ads so that the users come back on an e-commerce website they have previously visited. It comes bundled with multiple bandit agents and use cases, including the effect of selection bias on offline agents, and stochasticity in user response and evolution, as well as in observed returning time. Its ease of configurability and experimentation makes it a desirable choice, but it does not address multi-step reasoning, slate recommendation and presentation bias.

\noindent\textbf{MARS-Gym~\citep{mars-gym}} aims to simulate online marketplaces, and is based on real data from such platforms. It includes next-item prediction and off-policy metrics for evaluation of trained agents. The agent's objective, for a given user, is to select one of the items that were observed in the real data for that user. Therefore, MARS-Gym aims to evaluate the quality of static, semantic information learned by agents and does not meet our research agenda targeting dynamic and interactive systems.

\noindent\textbf{RL4RS~\citep{rl4rs}} is an e-commerce, slate recommendation simulator based on real purchase data, and where the reward function is a black-box sequential recommendation model. It is composed of two variants: one-shot (i.e., single-turn) and sequential slate recommendation. Offline RL agents can be trained on the real logged data and evaluated in the simulator, but one cannot directly control the logging policy, and presentation bias is not modeled. The authors verify that a transformer model can better capture the item sequence using non-greedy decoding strategies, which might indicate multi-step dependencies. However, the simulator is opaque and hardly tunable, and thus does not satisfy our specifications for a research-oriented simulator.

\noindent\textbf{RecSim~\citep{recsim}} is certainly the effort closest to ours. The authors provided a configurable simulator and three environment instantiations that cover, at least partially, all research topics that we wish to study. However, we found practical drawbacks in using it, motivating us to propose our take on interactive recommendation simulators. First, installation and usage is made very difficult as it relies on older, unmaintained packages, without specifying the version being used. Moreover, relying on third-party software like Tensorflow 1.15 or Google dopamine hurts the ease of configurability of both the environment and agents. In contrast, our simulator relies only on Numpy (and Gymnasium). Second, we found tweaking the environment properties and singling out specific research questions to be hard, as there are often multiple parameters controlling the same research dimension, without clear guidance on their effect, and they are not always tunable without substantial modifications: e.g., the uncertainty in user response comes from a binary coin flip, which does not allow to draw profiles of robustness to increasing uncertainty. There is also no simple way to use an oracle for user state or item embeddings as we do in our environment to single out certain modules of the agent. Finally, while the simulator aims to tackle slate recommendation, no proposed environment uses a number of candidates greater than $15$ or a slate size greater than $3$, while we wish to study slate recommendation at a larger scale, e.g., with a number of candidates of $1000$ and a slate size of $10$ as in our proposed \texttt{SlateTopK} environments. Overall, we adopt the general philosophy of RecSim and propose our take on making a lightweight, flexible, and research-oriented simulator.

\noindent\textbf{Virtual-TaoBao~\citep{virtual-tb}} is an online retail simulator trained from real data, where generative adversarial networks are trained via multi-agent imitation learning in order to approximate the user response to recommendations. It incorporates certain uncertainties, e.g., on the user churning mechanism, and rewards multi-step reasoning, but it does not address other research topics, i.e., biases in the data and slate recommendation. Additionally, since the simulator consists of model approximations of real user behavior, the notion of items is lost (state and actions are continuous latent variables) and the user response is a black-box that cannot be tweaked for further experimentations.

\noindent\textbf{SOFA~\citep{sofa}} uses an intermediate reweighting step in order to remove popularity and positivity biases in the resulting simulator. The authors verify that policies trained in the debiased simulators perform better when evaluated on datasets from the same platform but where biases have been alleviated (i.e., through random recommendations). The simulator is relatively easy to tweak as we can replace the intermediate inverse-propensity scoring step with a different technique, and change the underlying logged data. However, SOFA does not target the study of the other research topics in our agenda, i.e., multi-step reasoning, environment uncertainty and slate recommendation.

\noindent\textbf{OBP~\citep{obp}} is a semi-synthetic, research-oriented simulator for off-policy training evaluation of bandit agents. Using real logs of an online retail platforms collecting with several policies, it can evaluate the quality of off-policy evaluation estimators and therefore help research in that direction. However, it does not address our other concerns, i.e., multi-step reasoning, environment uncertainty and slate recommendation.

\section{Conclusion}
\label{conclusion}

\paragraph{\bf Summary.} In this paper, we have introduced SARDINE, a simulator for automated recommendation in a dynamic and interactive environments. Our efforts seek to address different shortcomings identified in existing recommendation simulators: (i) a lack of comprehensiveness in the covered research questions, that compels researchers and practioners to scatter their study across several simulators; (ii) a lack of interpretability and controllability, when specific aspects of the simulator depend on the setting of multiple parameters; (iii) the inability to study in isolation the phenomena and effects of interest in the simulator; (iv) the solvability of the simulator through trivial off-the-shelf baselines; and (v) the difficulty for researchers and practitioners to make additions and changes to the simulator to study certain directions in more depth, or to investigate new research questions.

In an effort to cover a wide range of problems studied in recommendation, we devised our simulator to enable the investigation of four over-arching research topics, including the multi-step reasoning capacity of models (RT1), the ability to learn models from biased data (RT2), the robustness to uncertainty (RT3), and the challenges associated with recommending slates (RT4). Concretely, these research topics translate into six dimensions~-- that the practioner may or may not decide to include in their instantiation of the simulator~-- spanning the recommendation type (single-item vs.\ slate recommendation), the inclusion of a boredom and/or influence mechanism, the level of uncertainty in the clicking process, the state observability (full vs.\ partial), and whether the task is reranking.

We then conducted extensive experiments on a set of nine environments derived from SARDINE. These environments have been selected to constitute diverse combinations of the aforementioned dimensions and thus provide a good coverage of our four research topics. In our experiments, we compared various methods which include both simple baselines and state-of-the-art approaches. To foster reproducibility and enable researchers to draw from this work to develop their own environments, we release the code for the simulator,\footnote{\href{https://github.com/naver/sardine}{https://github.com/naver/sardine}} as well as the detailed specifications for each environment and the implementation for all the compared methods.\footnote{\href{https://github.com/RomDeffayet/SARDINE\_Experiments}{https://github.com/RomDeffayet/SARDINE\_Experiments}}

\paragraph{\bf Findings.} Through our experiments on nine \acs{SARDINE} environments, we derived some valuable insights on the behavior of existing approaches in certain settings, demonstrating the usefulness of our simulator. First, we found that the SAC + Top-K approach, which combines the widely used SAC agent to a simple top-K ranker, showed impressive performance across the different environments and demonstrated a high stability. To the best of our knowledge, this approach is rarely considered as a baseline in RL-based slate recommendation works (except in \citep{gems}) despite its effectiveness and relative simplicity in comparison to state-of-the-art models. Therefore, we advocate for its usage as a baseline in future work on slate recommendation in dynamic environments.

To slightly nuance this first finding, we wish to add as a caveat that SAC + Top-K may be particularly dependent on the high quality of the item embeddings used. The performance of this approach was particularly high when using the ideal item embeddings (i.e., the ones that are used inside the simulator), but it decreased by a good margin when we replaced the ideal item embeddings with sub-optimal, matrix factorization embeddings. In comparison, the SAC + GeMS \citep{gems} approach seemed to be more robust overall to the item embedding quality. The recent \acf{HAC} approach \citep{Liu2023} was the most impacted by the quality of item embeddings in the studied settings. Moreover, we found that this approach was more affected than other methods by a highly dynamic environment with a drift in the user interests. We attribute this to the supervised click prediction loss used in HAC, which favors immediate reward over multi-step reasoning in the model.

Secondly, we studied how a transformer state encoder compare to a GRU state encoder in partially observable environments, and identified that the former tends to outperform the latter. This was notable in particular on environments where the click uncertainty was medium or high. This finding on the superiority of the transformer over the GRU as a state encoder goes in line with previous studies~\citep{Huang2022} and thus it does not come as a surprise. However, the impact of the level of click uncertainty on the state encoder is a subject that has not been considered a lot in the recommendation literature, and might be a topic worth investigating more deeply. 

Finally, we conducted experiments on the impact of presentation bias in the user feedback in a recommendation scenario. We notably found that when the environment is dynamic, click models trained offline may be less accurate than on static environments, which can have a detrimental effect on downstream tasks, such as counterfactual learning-to-rank or offline reinforcement learning. Our experiments also open up the possibility of studying the end-to-end training of RL agents from biased data, including a click modeling step.

Overall, the experiments we conducted act as guidance on how to use the simulator, examples of use cases that can be studied through \acs{SARDINE}, and more importantly as a call for more research on dynamic and interactive approaches for recommender systems. The insights we gathered throughout our experiments also reinforce the usefulness of simulated evaluation in general, and \acs{SARDINE} in particular.

\paragraph{\bf Future work.} While we designed our simulator to be flexible and configurable, so that researchers can tweak the experimental setup to their needs, we did not implement variants of the simulator that target the study of many of the research questions associated with our agenda~(Section~\ref{sec:research-agenda}). For instance, the performance of agents when the environment is non-stationary (e.g., due to changes in the world) is still largely unknown~\citep{non-stationary-RL}, and could be investigated in \acs{SARDINE}. Similarly, reaching the best possible policy in a limited number of deployments, a task known as deployment-efficiency~\citep{deployment-efficiency}, as well as continual learning~\citep{continual-learning}, which aims to deploy agents that keep on learning, could be explored in the recommendation scenario thanks to \acs{SARDINE}. We hope our simulator can foster the experimentation of such novel ideas in recommender systems research.

\begin{acks}
This research was (partially) funded by 
the Hybrid Intelligence Center, a 10-year program funded by the Dutch Ministry of Education, Culture and Science through the Netherlands Organisation for Scientific Research, \url{https://hybrid-intelligence-centre.nl}, 
project LESSEN with project number NWA.1389.20.183 of the research program NWA ORC 2020/21, which is (partly) financed by the Dutch Research Council (NWO),
and
the FINDHR (Fairness and Intersectional Non-Discrimination in Human Recommendation) project that received funding from the European Union’s Horizon Europe research and innovation program under grant agreement No 101070212.

All content represents the opinion of the authors, which is not necessarily shared or endorsed by their respective employers and/or sponsors.
\end{acks}

\appendix
\section{Efficiency}
\label{app:efficiency}

On a single Intel Xeon Gold 6338 CPU, we found that our simulator can operate at approximately $4,500$ steps (i.e., user interactions) per second with the \texttt{SlateTopK-Bored} environment and up to $5,000$ steps per second on the \texttt{SingleItem-Static} environment. Moreover, training a SAC+TopK agent on \texttt{SlateTopK-Bored} or \texttt{SlateTopK-BoredInf} for $500,000$ training steps, as in Section~\ref{sec:exp-topk}, takes around $40$ minutes on a single NVIDIA A100 GPU. 

\section{Webtoon experiment}
\label{app:webtoon}

\subsection{Environment description}

The environments introduced in Section~\ref{experimental-setup} and experimented on in Section~\ref{experiments} are based on purely synthetic items with uniformly drawn topic-item assignments. While these environments enabled us to derive interesting insights, one might question whether such environments reflect a real-life scenario where (i) topic-item assignments are not uniformly drawn (i.e., some topics tend to co-occur within items), (ii) item-topic distribution is skewed (i.e., some topics are more prominent than others among items), and (iii) user-topic distribution is skewed (i.e., some topics are more popular than others among users). To showcase the possibilities of SARDINE to address such a scenario, we define a semi-synthetic environment named \texttt{WebtoonSlateTopK-Bored} that presents the same characteristics as \texttt{SlateTopK-Bored} with the difference that its items are based on the real-world catalog of the Webtoon\footnote{\url{https://www.webtoons.com/en} (item catalog accessed in January 2022).} online comics platform. The hyperparameters for this environment are summarized in Table~\ref{tab:hyperparam-val-webtoon}. Differently from \texttt{SlateTopK-Bored}, \texttt{WebtoonSlateTopK-Bored} includes 669 items and 16 topics.

\begin{table}[t]
\centering
\caption{Value of the simulator hyperparameters for the \texttt{WebtoonSlateTopK-Bored} environment. The description of the hyperparameters' meaning and role is detailed in Table~\ref{tab:hyperparam-desc}.}
\vspace{0.2cm}
\label{tab:hyperparam-val-webtoon}
\setlength{\tabcolsep}{1.4mm}
\begin{tabular}{l cccccccccccc l}
\toprule
\multirow{2}{*}{Environment name} & \multicolumn{12}{c}{Hyperparameter value} &  \\ \cmidrule(l){2-14} 
 & $L$ & $S$ & $n_{\mathcal{I}}$ & $n_{\mathcal{T}}$ & $\lambda$ & $\mu$ & $\alpha$ & $\epsilon$ & $n_{\text{b}}$ & $t_{\text{b}}$ & $\tau_{\text{b}}$ & $\omega$ & $\mathcal{O}$ \\ \midrule
\texttt{WebtoonSlateTopK-Bored} & 100 & 10 & 669 & 16 & 100 & 0.65 & 1.0 & 0.85 & 10 & 5 & 5 & 1.0 & full \\ 
\bottomrule
\end{tabular}
\end{table}

\begin{figure*}[t]
\centering
    \begin{subfigure}{.495\textwidth}
    \includegraphics[width = \textwidth]{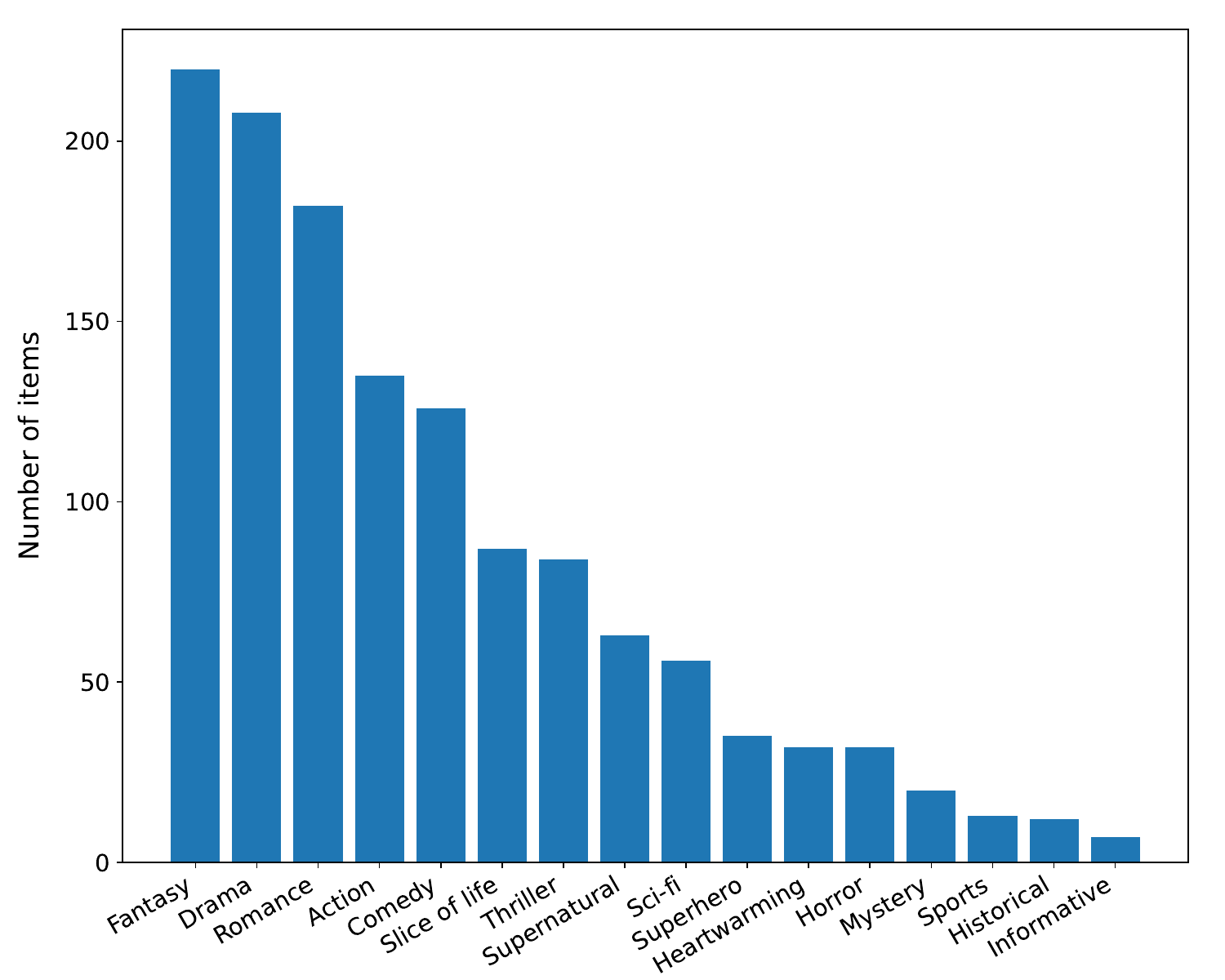} 
    \caption{Item distribution across topics}
    \label{fig:webtoon-items-per-topic}
    \end{subfigure}
    \hfill
    \begin{subfigure}{.495\textwidth}
    \includegraphics[width = \textwidth]{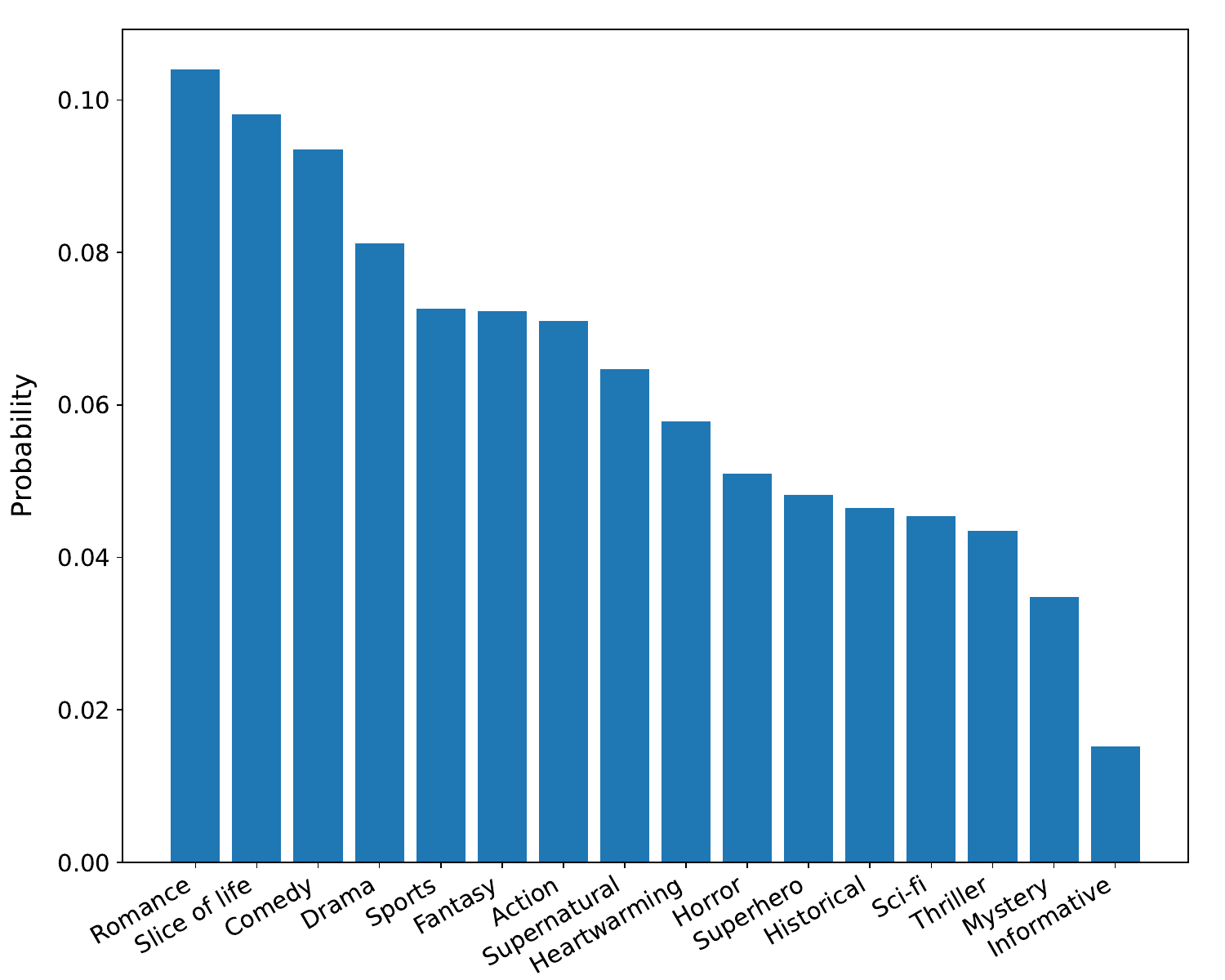} 
    \caption{Topic prior for user embeddings}
    \label{fig:webtoon-topic-prior}
    \end{subfigure}
\vspace{0.1cm}
\caption{Properties of Webtoon items and topics (i.e., Webtoon categories). Fig. \ref{fig:webtoon-items-per-topic} depicts the number of items attributed to each topic, based on the actual item-category assignments in the Webtoon catalog. Fig. \ref{fig:webtoon-items-per-topic} illustrates the topic prior used for generating user embeddings in SARDINE, which is drawn from the number of likes obtained for the items of each category.}
\label{fig:webtoon-properties}       
\end{figure*}

\paragraph{\bf User and item embeddings.} In this experiment, we consider that we only have access to an item catalog, which does not directly include embeddings. Therefore, item and user embeddings need to be generated under the constraints imposed by the catalog (e.g., item-topic assignments), leading to a semi-synthetic setting. To obtain item and user embeddings based on the Webtoon catalog, we slightly changed the generation process described in Section~\ref{sec:embeddings}. For item embeddings, steps (2) and (3) are removed as item-topic assignments are directly obtained from the catalog~-- these correspond to the genres of an item, such as \textit{Drama}, \textit{Romance}, \textit{Superhero}, \textit{Sci-fi}, etc. Exploiting these assignments ensures a meaningful co-occurrence of topics within items: for example, the pairs (\textit{Drama}, \textit{Romance}) as well as (\textit{Superhero}, \textit{Sci-fi}) are more likely to co-occur than the pair (\textit{Romance}, \textit{Superhero}). The distribution of items across topics, i.e., the number of items attributed to each topic, is illustrated in Fig.~\ref{fig:webtoon-items-per-topic}. This figure highlights a skewed distribution, with a large share of items pertaining to \textit{Fantasy}, \textit{Drama}, or \textit{Romance} topics, and a much smaller share of items with \textit{Sports}, \textit{Historical} or \textit{Informative} topics.

To generate user embeddings, we changed step (3) from the embedding generation process in Section~\ref{sec:embeddings} to reflect the fact that topics may not be uniformly popular among users. More specifically, instead of being sampled from $\text{Unif}(\mathcal{T})$, the topics of interest for a user $u$ (denoted as $\mathcal{T}_u = \{T_{u,1}, \ldots, T_{u,n_{\mathcal{T}_u}}\} \subset \mathcal{T}$) are sampled from a categorical, non-uniform prior $p_{\mathcal{T}}$ without replacement. The probability $p_{\mathcal{T}}(j)$ for a topic $j$ is defined as the ratio of the average number of likes for items with category $j$ divided by the average number of likes for items with any category. Formally, $p_{\mathcal{T}}(j)$ is defined as follows:
\begin{equation}
    p_{\mathcal{T}}(j) = \frac{\frac{1}{|\mathcal{I}_j|} \sum_{i \in \mathcal{I}_j} \mathrm{\#likes}(i)}{\sum_{j' \in \mathcal{T}} \frac{1}{|\mathcal{I}_{j'}|} \sum_{i \in \mathcal{I}_{j'}} \mathrm{\#likes}(i)},
\end{equation}
where $\mathrm{\#likes}(i)$ indicates the number of likes given to item $i$, and $\mathcal{I}_j$ is the set of items which pertain to topic $j$ (i.e., items $i$ such that $T_{i,j} > 0$). The distribution $p_{\mathcal{T}}$ is plotted in Fig.~\ref{fig:webtoon-topic-prior}, which highlights as well the skewed nature of user-topic preferences in the \texttt{WebtoonSlateTopK-Bored} environment.

\subsection{Results}

\begin{figure*}[t]
\centering
    \begin{subfigure}{.495\textwidth}
    \includegraphics[width = \textwidth]{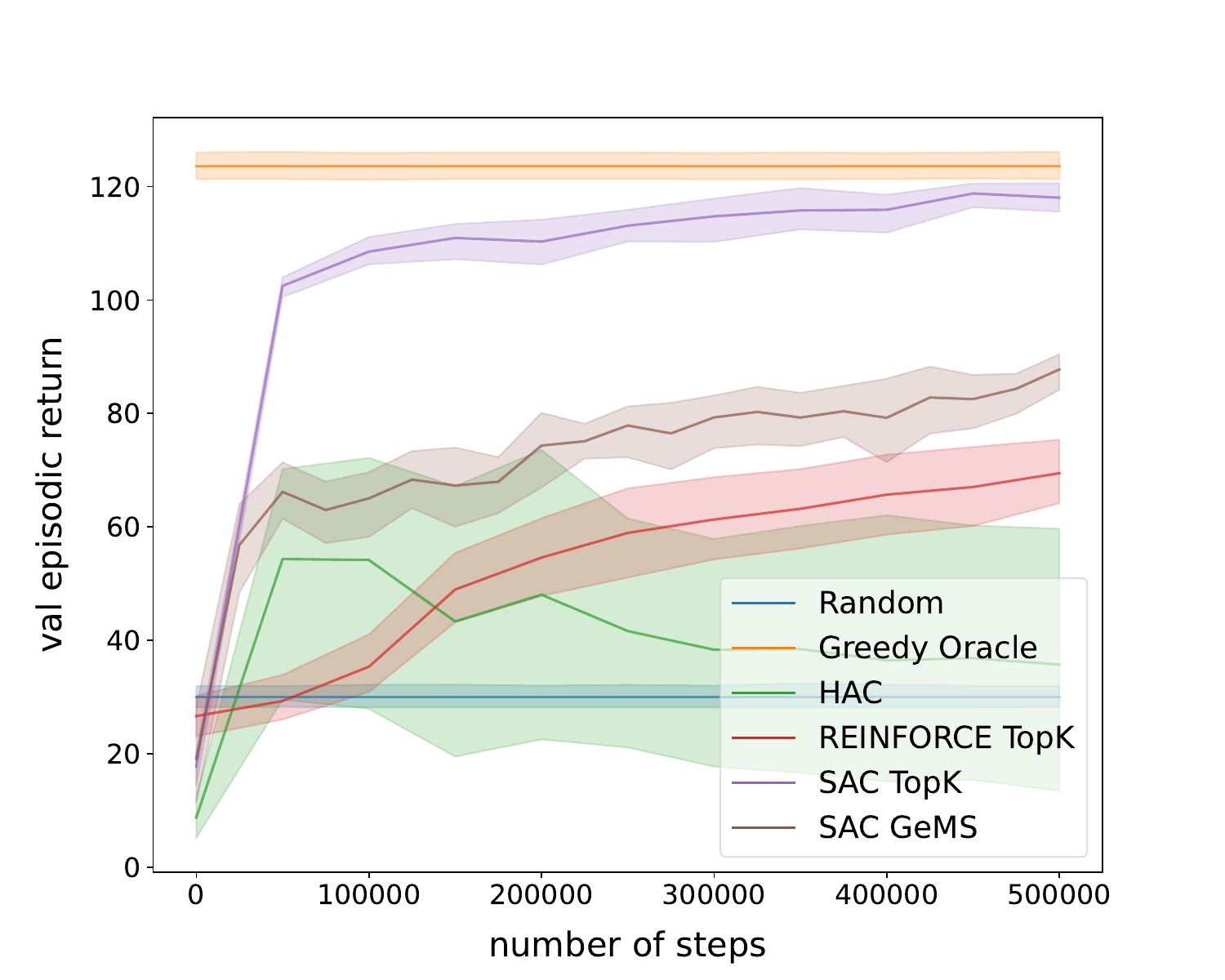} 
    \caption{Return ($\uparrow$)}
    \label{fig:WebtoonSlateTopK-Bored-return}
    \end{subfigure}
    \hfill
    \begin{subfigure}{.495\textwidth}
    \includegraphics[width = \textwidth]{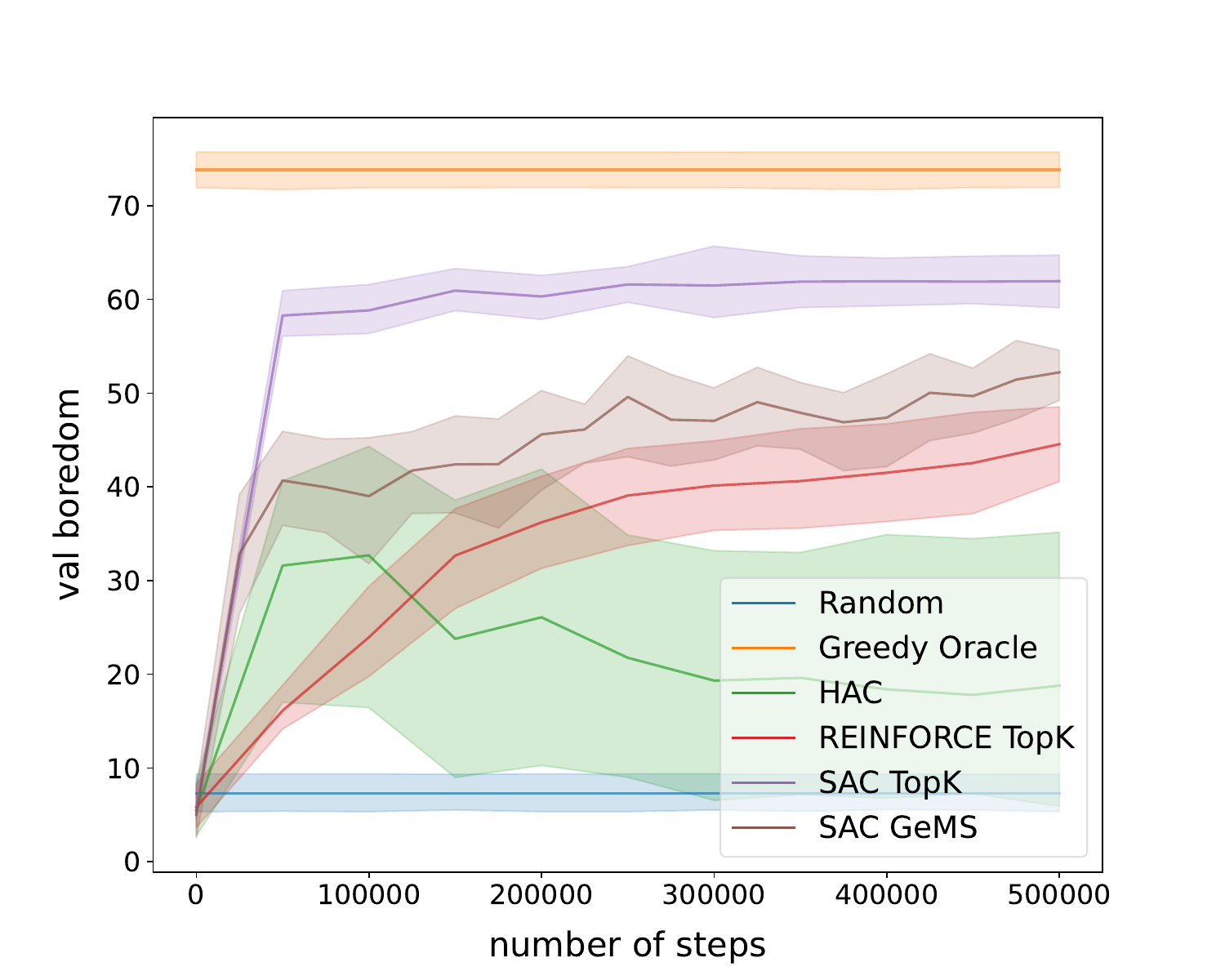} 
    \caption{Boredom ($\downarrow$)}
    \label{fig:WebtoonSlateTopK-Bored-boredom}
    \end{subfigure}
\vspace{0.1cm}
\caption{Results on the \texttt{WebtoonSlateTopK-Bored} environment. The colored envelope surrounding lines indicates the 95\% confidence interval around the mean computed from 5 seeded runs.}
\label{fig:WebtoonSlateTopK}       
\end{figure*}

Following the protocol and hyperparameters used for the experiments on \texttt{SlateTopK-Bored} (Ideal) (the results of which are described in Section~\ref{sec:exp-topk}, and illustrated in Fig.~\ref{fig:SlateTopK-Bored-return} and Fig.~\ref{fig:SlateTopK-Bored-boredom}), we compared the same methods on the \texttt{WebtoonSlateTopK-Bored} environment. The results are shown in Fig.~\ref{fig:WebtoonSlateTopK}, with the return in Fig.~\ref{fig:WebtoonSlateTopK-Bored-return} and the boredom in Fig.~\ref{fig:WebtoonSlateTopK-Bored-boredom} averaged over validation episodes. A notable difference with the results on the \texttt{SlateTopK-Bored} environment is the fact that no RL-based approach is able to beat the Greedy Oracle. This might be explained by the greater difficulty of \texttt{WebtoonSlateTopK-Bored} due to its skewed item-topic and user-topic distributions. Nonetheless, similarly to the results on \texttt{SlateTopK-Bored}, we observe that among the RL-based approaches, SAC + Top-K performed the best and is followed by SAC + GeMS. However, differently from previous results, HAC underperformed and showed some instability which is illustrated by its high variance. This suggests that HAC might be less robust and more sensitive to changes in the environment in comparison to other methods. Overall, this experiment demonstrates that deriving environments with realistic, long-tail distributions provides interesting and challenging use-cases in SARDINE.

\begin{figure}[h]
\centering
\includegraphics[width = .7\textwidth]{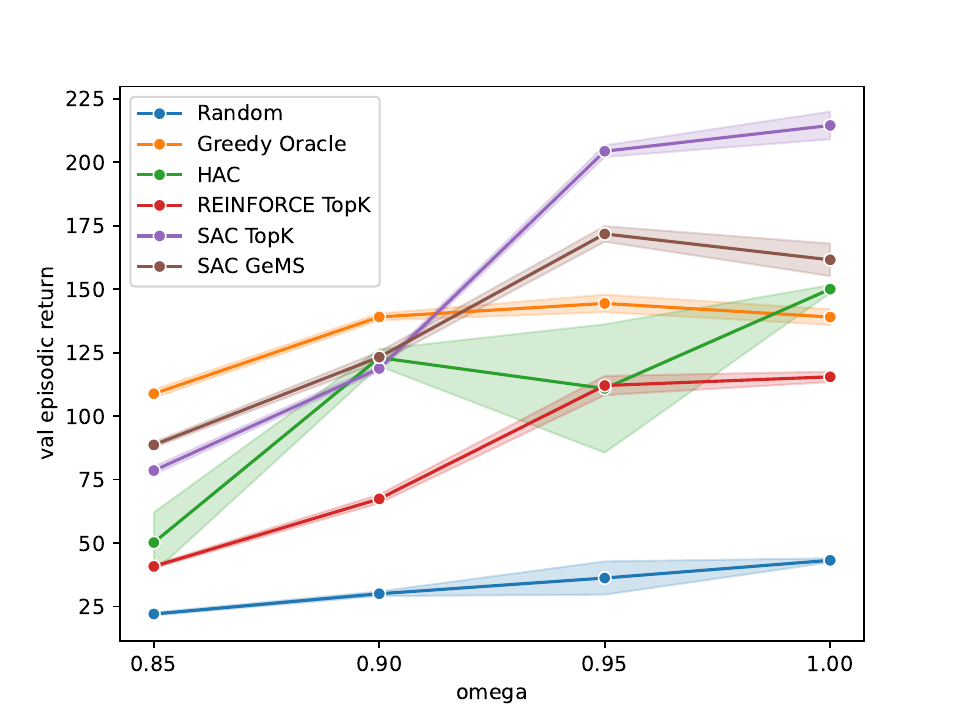} 
\vspace{0.1cm}
\caption{Validation return after $500,000$ training steps on the \texttt{SlateTopK-BoredInf} environment at different levels of influence by the clicked items (the lower $\omega$, the higher the influence of clicked items on user preference). The colored envelope surrounding lines indicates the 95\% confidence interval around the mean computed from 5 seeded runs.}
\label{fig:omega_comparison}   
\end{figure}

\section{Clicked item influence experiment}
\label{app:omega}

In this section we focus on the effect of clicked items on the user preference, which can be controlled by the $\omega$ parameter introduced in Section~\ref{sec:boredom}. Concretely, we use the \texttt{SlateTopK-BoredInf} environment and build a performance profile of the benchmarked algorithms under increasing influence of the clicked items: $\omega \in \{ 1.0, 0.95, 0.9, 0.85 \}$. Note that $\omega = 1$ correponds to the \texttt{SlateTopK-Bored} environment, where clicked items do not change the user preference. With $\omega < 1$, clicked items attract the user by shifting the user preference embedding towards the embedding of the clicked item. As a result, the system must control for the long-term effect of its recommendations on user preference, possibly yielding complex dynamics.

As in Section~\ref{sec:exp-topk}, we train all models for $500,000$ steps, with five different random seeds. In Figure~\ref{fig:omega_comparison}, we report the validation episodic return obtained after $500,000$ training steps, at four different values of $\omega$. We find that the complex dynamics created by increasing the influence of clicked items makes it harder for all methods to reach a high return. In particular, for $\omega \leqslant 0.9$, none of the tested methods manages to beat a greedy oracle agent. Moreover, the relative performance of the different methods is sensitive to the strength of item influence, with SAC+GeMS being overall more robust than its counterparts and even beating SAC+TopK under lower $\omega$ values.

\section{Item scores}
\label{app:scores}

\begin{figure}[h]
\centering
\includegraphics[width = .7\textwidth]{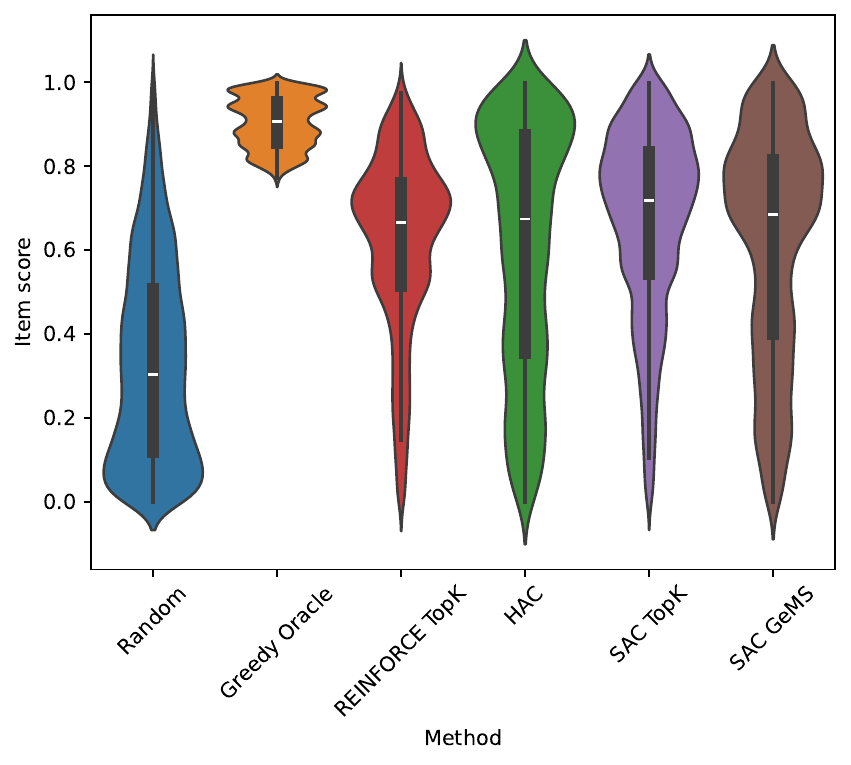} 
\caption{Distribution of the relevance score of items recommended by different methods after training for $500,000$ steps on \texttt{SlateTopK-Bored}. The higher the score, the higher the click probability. See Section~\ref{sec:click} for how the relevance score is computed within the simulator.}
\label{fig:item-scores}   
\end{figure}

In Figure~\ref{fig:item-scores} we report the distribution of scores for items recommended by the methods benchmarked on the \texttt{SlateTopK-Bored} environment. In this environment, there exists a trade-off between recommending highly relevant items and mitigating user boredom. We can indeed see that while the greedy oracle algorithm recommends only highly relevant items, it does not yield as much return as multi-step approaches like HAC, SAC+TopK and SAC+GeMS. Furthermore, we can spot differences across methods, as it appears that HAC often recommends highly relevant items, but also often defaults to poorly relevant items, in contrast to REINFORCE+TopK, which mostly avoids irrelevant items but also usually fails to accurately estimate user preferences and propose highly relevant items.

Overall, the availability and interpretability of standardized item scores within our simulator\footnote{Item scores are returned by the \texttt{step} method of the simulator under \texttt{info["scores"]}.} allows us to complement the information contained in the final return obtained by each method, and therefore to better qualify the strengths and weaknesses of each method.

\clearpage
\bibliographystyle{ACM-Reference-Format}
\bibliography{main}

\end{document}